\documentclass[twocolumn,showpacs,amsmath,amssymb,floatfix]{revtex4}
\usepackage{citesort}
\usepackage{graphicx}
\usepackage{dcolumn}

\newcommand{\ecoli}{{\it E.~coli}}
\newcommand{\lacrep}{{\it lac} repressor}
\newcommand{\Lacrep}{{\it Lac} repressor}

\newcommand{\lr}{{\it lac} repressor}
\newcommand{\Lr}{{\it Lac} repressor}
\newcommand{\lo}{{\it lac} operon}

\newcommand{\Rmsd}{R.m.s.d.}
\newcommand{\rmsd}{r.m.s.d.}

\newcommand{\rv}{\vec{r}}
\newcommand{\qi}{q_{i=1-4}}

\newcommand{\dg}{^\circ}
\newcommand{\dgbp}{^\circ/{\rm bp}}
\newcommand{\dbp}{\mbox{\scriptsize deg/bp}}

\newcommand{\kT}{{\rm kT}}

\newcommand{\halfof}[1]{\frac{#1}{2}}

\newcommand{\di}{\vec{d}_i}
\newcommand{\dv}{\vec{d}_1}
\newcommand{\dt}{\vec{d}_2}
\newcommand{\dth}{\vec{d}_3}
\newcommand{\dframe}{$(\vec{d}_1, \vec{d}_2, \vec{d}_3)$}
\newcommand{\dframes}[1]{$(\:\vec{d}_1(#1), \;\vec{d}_2(#1), \;\vec{d}_3(#1)\:)$}

\newcommand{\ko}{\kappa_1}
\newcommand{\kt}{\kappa_2}
\newcommand{\kot}{\kappa_{1,2}}

\newcommand{\kiot}{\kappa_{1,2}^\circ}

\newcommand{\omo}{\omega^\circ}
\newcommand{\psio}{\psi_\circ}
\newcommand{\Co}{C_\circ}

\newcommand{\QD}{Q_{\mbox{\tiny DNA}}}
\newcommand{\ND}{N_{\mbox{\tiny DNA}}}

\newcommand{\E}{\vec{E}}
\newcommand{\wE}{w_{\mbox{\tiny {\it E}}}}

\newcommand{\refgrand}{(\ref{eq:grand_1}-\ref{eq:grand_11})}
\newcommand{\refgrandA}{(\ref{eq:adv_grand_1}-\ref{eq:adv_grand_11})}

\newcommand{\UP}{U$^\prime$}
\newcommand{\OP}{O$^\prime$}
\newcommand{\sP}{s^\prime}

\newcommand{\veps}{\vec{\epsilon}}
 
\begin{document}

 
\title{Modeling DNA loops using the theory of elasticity}

\author{Alexander Balaeff}
 \altaffiliation[Also ]{Center for Biophysics and Computational Biology, University of Illinois at Urbana-Champaign}%
\author{L. Mahadevan}
\altaffiliation{%
Department of Applied Mathematics and Theoretical Physics, %
University of Cambridge, Silver St., Cambridge, CB3 9EW, U.K.}%
\author{Klaus Schulten}%
 \altaffiliation[Also ]{Department of Physics, University of Illinois at Urbana-Champaign}%
 \email{kschulte@ks.uiuc.edu}
\affiliation{%
Beckman Institute,\\
University of Illinois at Urbana-Champaign,\\
Urbana, IL 61801}%

\date{\today}

\begin{abstract}
A versatile approach to modeling the conformations and energetics of DNA loops is presented. The model is based on the classical theory of elasticity, modified to describe the intrinsic twist and curvature of DNA, the DNA bending anisotropy, and electrostatic properties. All the model parameters are considered to be functions of the loop arclength, so that the DNA sequence-specific properties can be modeled. The model is applied to the test case study of a DNA loop clamped by the {\it lac} repressor protein. Several topologically different conformations are predicted for various lengths of the loop. The dependence of the predicted conformations on the parameters of the problem is systematically investigated. Extensions of the presented model and the scope of the model's applicability, including multi-scale simulations of protein-DNA complexes and building all-atom structures on the basis of the model, are discussed.
\end{abstract}


\maketitle
 

\section{Introduction}
 
Formation of DNA loops is a common motif of protein-DNA interactions~\cite{BERG2002,ALBE2002,MATT92,SCHL92A}.  A segment of DNA forms a loop-like structure when either its ends get bound by the same protein molecule or a multi-protein complex, or when the segment gets wound around a large multi-protein aggregate, or when the segment connects two such aggregates. In bacterial genomes, DNA loops were shown to play important roles in gene regulation~\cite{MATT92,SCHL92A}; in eucaryotic genomes, DNA loops are a common structural element of the condensed protein-DNA media inside nuclei~\cite{BERG2002,ALBE2002}.  Understanding of the structure and dynamics of DNA loops is thus vital for studying the organization and function of the genomes of living cells.

The amount of experimental data on the DNA physical properties and protein-DNA interactions both {\it in vivo} and {\it in vitro} has grown dramatically in recent years~\cite{MAHE2002}. With the advent of modern experimental techniques, such as micromanipulation~\cite{WILL2002,STRI2000,YIP2002} and fast resonance energy transfer~\cite{TRUO2002}, researchers were presented with unique opportunities to probe the properties of individual macromolecules. X-ray crystallography, NMR, and 3D electron cryomicroscopy~\cite{UNGE2002} provided numerous structures of protein-DNA complexes with resolution up to  a few angstroms, including such huge biomolecular aggregates as RNA polymerase~\cite{CRAM2002} and nucleosome~\cite{LUGE97,AREN91}.  The ever growing volume of data provides theoretical modeling, which has generally been recognized as a vital complement of experimental studies, with an opportunity to revise the existing models, and to build new improved models of biomolecules and biomolecular interactions.

Several existing DNA models are based on the theory of elasticity~\cite{OLSO2000,OLSO96,SCHL95,VOLO94}.  These models treat DNA as an elastic rod or ribbon, sometimes carrying an electric charge.  The geometrical, energetic, and dynamical properties of such ribbon can be studied at finite temperature using Monte-Carlo or Brownian Dynamics techniques employing a combined elastic/electrostatic energy functional~\cite{OLSO2000,OLSO96}. Such studies usually involve extensive data generation, for example, numerous Monte Carlo structural ensembles, and require significant investment of computational resources.  Alternatively, one could resort to faster theoretical methods, such as statistical mechanical analysis of the elastic energy functional~\cite{MARK95A,MARK98} or normal mode analysis of the dynamical properties of the elastic rod~\cite{MATS2002}. A fast approach to studying the static properties of DNA loops -- such as the loop energy, structure, and topology -- consists in solving the classical equations of elasticity~\cite{LAND86,LOVE27}, dated back to Kirchhoff~\cite{KIRCH1876} and derived on the basis of the same energy functional. The equations can be solved with either fixed boundary conditions for the ends of the loop or under a condition of a constant external force acting on the ends of the loop~\cite{TOBI2000,COLE2000,WEST97,COLE95,SHI95,SHI94}.

In order to achieve a realistic description of the physical properties of DNA, the classical elastic functional~\cite{LAND86,LOVE27,MAHA96} has to be modified: (i) the modeled elastic ribbon has to be considered intrinsically twisted (and, possibly, intrinsically bent) in order to mimic the helicity of DNA, (ii) the ribbon has to carry electrostatic charge, (iii) be anisotropically flexible, i.e., different bending penalties should be imposed for bending in different directions, (iv) be deformable, e.g., through extension and/or shear, (v) have different flexibility at different points, in order to account for DNA sequence-specific properties, (vi) be subject to possible external forces, such as those from proteins or other DNA loops. The earlier works have presented many models including several of these properties, e.g., bending anisotropy and sequence-specificity~\cite{MATS2002}; extensibility, intrinsic twist/curvature, and electrostatics~\cite{WEST97}; extensibility, shearability, and intrinsic twist~\cite{SHI95}; intrinsic curvature and electrostatics~\cite{KATR97}; or intrinsic twist and forces due to self-contact~\cite{TOBI2000,COLE2000}.

Yet, to the best of our knowledge, a completely realistic treatment, where the theory of elasticity would be modified as to include all of the listed DNA properties, has never been published. While for DNA segments of large length some of these properties can be disregarded or averaged out using a proper set of effective parameters, we feel that a proper model of DNA on the scale of several hundred base pairs -- which is a typical size of DNA loops involved in protein-DNA interactions -- must be detailed, including a proper description of all physical properties of real DNA.

This work offers a step towards such a generalized elastic DNA model. The Kirchhoff equations of elasticity are derived in Sec.~\ref{sec:theory} below for an intrinsically twisted (and possibly bent) elastic ribbon with anisotropic bending properties. The terms corresponding to external forces and torques are included and can also be used to account for the electrostatic self-repulsion of the rod, as described in Sec.~\ref{sec:electrostatics}.  All the parameters are considered to be functions of the ribbon arclength $s$, thus making the DNA model sequence-specific. Only the DNA deformability is omitted from the derived equations - yet can be straightforwardly included in the problem, as discussed in Sec.~\ref{sec:discussion}. The numerical algorithm for solving the modified Kirchhoff equations, based on the earlier work~\cite{MAHA96,BALA99}, is presented (Sec.~\ref{sec:lacsols_short}).

The proposed model is used to predict and analyze the structure of the DNA loops clamped by the {\it lac} repressor, a celebrated {\it E.~coli} protein, reviewed in Sec.~\ref{sec:lac_operon}. The system provides a typical biological application for the developed model and is used to extensively analyze the effect of bending anisotropy (Sec.~\ref{sec:anisotropy}) and electrostatic repulsion (Sec.~\ref{sec:electrostatics}) in our model and to evaluate the corresponding model parameters. The DNA sequence-specific properties in terms of elastic moduli and intrinsic curvature, while included in the derived equations, were not used in the study of the \lr\ system.

The further developments that would make the model truly universal and the model's scope of applicability are discussed in Sec.~\ref{sec:discussion}. Notably, it is shown how the elastic rod model can be combined with the all-atom model in multi-scale simulations of protein-DNA complexes or how all-atom DNA structures can be built on the basis of the coarse-grained model.
 
 
\section{Basic Theory of Elasticity}
\label{sec:theory}

In this section we describe first the classical Kirchhoff theory of elasticity and then how this theory can be applied to modeling DNA loops.


\subsection{Kirchhoff model of an elastic rod}

\begin{figure}[!t]
\begin{center}
\includegraphics[scale=.55]{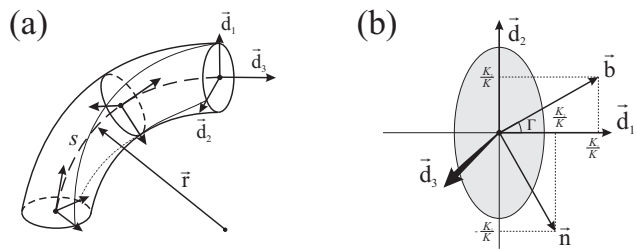}
\end{center}
\caption{\label{fig:elrod}Parameterization of the elastic rod. (a) The centerline $\rv(s)$ and the intrinsic local frame $\dv$, $\dt$, $\dth$. (b) The principal normal $\vec{n}(s) = \ddot{\rv} / |\ddot{\rv}|$ and the binormal $\vec{b}(s) = \dth \times \vec{n}$ form the natural local frame for the 3D curve $\rv(s)$~\cite{KREY91}.}
\end{figure}

The classical theory of elasticity describes an elastic rod (ribbon) in terms of its centerline and cross sections (Fig.~\ref{fig:elrod}\,a). The centerline forms a three-dimensional curve $\rv(s) = \left( x(s), y(s), z(s) \right)$ parametrized by the arclength~{\it s}.  The cross sections are ``stacked'' along the centerline; a frame of three unit vectors $\dv(s)$, $\dt(s)$, $\dth(s)$ uniquely defines the orientation of the cross section at each point~$s$.  The vectors $\dv$ and $\dt$ lie in the plane of the cross section (for example, along the principal axes of inertia of the cross section), and the vector $\dth = \dv \times \dt$ is the normal to that plane~\footnote{All the variables in this paper, apart from a few clear exceptions, are considered to be functions of arclength $s$.  Therefore, we frequently drop the explicit notation ``$(s)$'' from the equations throughout the paper.}.  If the elastic rod is inextensible then the tangent to the center line coincides with the normal $\dth$:
\begin{equation}
\label{eq:inext}
\dot{\rv}(s) = \dth(s)
\end{equation}
\noindent (the dot denotes the derivative with respect to $s$).

The components of all the three vectors $\di$ can be expressed through three Euler angles $\phi(s)$, $\psi(s)$, $\theta(s)$, which define the rotation of the local coordinate frame \dframe\ relative to the lab coordinate frame.  Alternatively, one can use four Euler parameters $q_1(s)$, $q_2(s)$, $q_3(s)$, $q_4(s)$, related to the Euler angles as
\begin{equation}
\begin{split}
\label{eq:eul_par}
q_1 & = \sin{\theta/2}\:\sin{( \phi - \psi )/2} , \\
q_2 & = \sin{\theta/2}\:\sin{( \phi - \psi )/2} , \\
q_3 & = \cos{\theta/2}\:\sin{( \phi + \psi )/2} , \\
q_4 & = \cos{\theta/2}\:\cos{( \phi + \psi )/2} ,
\end{split}
\end{equation}
\noindent and subject to the constraint
\begin{equation}
\label{eq:q_constr}
q_1^2 + q_2^2 + q_3^2 + q_4^2 = 1
\end{equation}
\noindent (see, e.g.,~\cite{WHIT60}). The computations presented in this paper employ the Euler parameters in order to avoid the polar singularities inherent in the Euler angles.

Following Kirchhoff's analogy between the sequence of cross-sections of the elastic rod and a motion of a rigid body, the arclength $s$ can be considered as a time-like variable. Then the spatial angular velocity~$\vec{k}$ of rotation of the local coordinate frame can be introduced:
\begin{equation}
\label{eq:fr_rot}
\dot{\vec{d}}_{i=1-3} = \vec{k} \times \di\;, \quad \vec{k} = \{ K_1, K_2, \Omega \} \;.
\end{equation}
\noindent The vector $\vec{k}$ is called the vector of strains~\cite{LAND86,LOVE27}. Geometrically, its components $K_1(s)$ and $K_2(s)$ are equal to the principal curvatures of the curve $\rv(s)$, so that the total curvature equals
\begin{equation}
\label{eq:tot_cur}
K(s) = \sqrt{K_1^2(s) + K_2^2(s)}\;,
\end{equation}
\noindent and the vectors of principal normal $\vec{n}(s)$ and binormal $\vec{b}(s)$ (Fig.~\ref{fig:elrod}\,b) are
\begin{eqnarray}
\label{eq:pr_nor}
\vec{n}(s) =& K_2/K\: \dv - K_1/K\: \dt \;,\\
\label{eq:binor}
\vec{b}(s) =& K_1/K\: \dv + K_2/K\: \dt \;.
\end{eqnarray}
\noindent The third component of the vector of strains $\Omega(s)$ is the local twist of the elastic rod around its axis $\dth$~\cite{LOVE27}.  All three components can be expressed via the Euler parameters $q_{i=1-4}$~\cite[p.16]{WHIT60}:
\begin{eqnarray}
\label{eq:cur_eu1}
K_1 & = & 2 ( q_4 \dot{q}_1 + q_3 \dot{q}_2 - q_2 \dot{q}_3 - q_1 \dot{q}_4 )\;, \\
\label{eq:cur_eu2}
K_2 & = & 2 ( - q_3 \dot{q}_1 + q_4 \dot{q}_2 + q_1 \dot{q}_3 - q_2 \dot{q}_4 )\;, \\
\label{eq:cur_euO}
\Omega & = & 2 ( q_2 \dot{q}_1 - q_1 \dot{q}_2 + q_4 \dot{q}_3 - q_3 \dot{q}_4 )\;.
\end{eqnarray}

If the elastic rod is forced to adopt a shape different from that of its natural (relaxed) shape, then elastic forces $\vec{N}(s)$ and torques $\vec{M}(s)$ develop inside the rod:
\begin{align}
\label{eq:for_torN}
\vec{N}(s) & = N_1 \dv + N_2 \dt + N_3 \dth\;, \\
\label{eq:for_torM}
\vec{M}(s) & = M_1 \dv + M_2 \dt + M_3 \dth\;.
\end{align}
\noindent The components $N_1$ and $N_2$ compose the shear force $N_{sh} = \sqrt{N_1^2+N_2^2}$; the component $N_3$ is the force of tension, if $N_3>0$, or compression, if $N_3<0$, at the cross section at the point $s$. $M_1$ and $M_2$ are the bending moments, and $M_3$ is the twisting moment.  In equilibrium, the elastic forces and torques are balanced at every point $s$ by the body forces $\vec{f}(s)$ and torques $\vec{g}(s)$, acting upon the rod:
\begin{gather}
\label{eq:for_eq}
\dot{\vec{N}} + \dot{\vec{f}} = 0\;, \\
\label{eq:tor_eq}
\dot{\vec{M}} + \dot{\vec{g}} + \dot{\rv} \times \vec{N} = 0\;.
\end{gather}
\noindent The body forces and torques of the classical theory usually result from gravity or from the weight of external bodies -- as in the case of construction beams.  In the case of DNA, such forces are mainly of electrostatic nature, as will be described below.

The last equation required to build a self-contained theory of the elastic rod relates the elastic stress to the distortions of the rod.  The classical approach stems from the Bernoulli-Euler theory of slender rods, which stipulates the elastic torques $\vec{M}(s)$ to be linearly dependent on the curvatures and twist of the inherently straight rod:
\begin{equation}
\label{eq:ber_eul}
\vec{M}(s) = A_1 K_1 \dv + A_2 K_2 \dt + C \Omega \dth \;.
\end{equation}
The linear coefficients $A_1$ and $A_2$ are called the bending rigidities of the elastic rod, and $C$ is called the twisting rigidity.  For a solid rod, the classical theory finds that $A_1 = E I_1$, $A_2 = E I_2$, where $E$ is the Young's modulus of the material of the rod, and $I_1$, $I_2$ are the principal moments of inertia of the rod's cross-section.  The twisting rigidity $C$ is proportional to the shear modulus $\mu$ of the material of the rod; in the simple case of a circular cross-section, $C = 2 \mu I_1 = 2 \mu I_2$.

The equations \eqref{eq:inext}, \eqref{eq:q_constr}, \eqref{eq:for_eq},  \eqref{eq:tor_eq}, and \eqref{eq:ber_eul} form the basis of the Kirchhoff theory of elastic rods.  We simplify the equations, first, by making all the variables dimensionless:
\begin{gather}
\bar{s} = s/l, \;\; \bar{x} = x/l, \;\; \bar{y} = y/l, \;\; \bar{z} = z/l\:, \\
\bar{K_1} = l K_1, \;\; \bar{K_2} = l K_2, \;\; \bar{\Omega} = l \Omega\:, \\
\label{eq:mod_scale_0}
\alpha = A_1/C, \;\; \beta = A_2/C, \;\; \gamma = C/C \equiv 1 \:, \\
\label{eq:for_scale_0}
\bar{N}_{i=1-3} = N_i l^2/C, \;\; \bar{M}_{i=1-3} = M_i l/C \:,
\end{gather}
\noindent where $l$ is the length of the rod. Second, we express the derivatives $\dot{q}_i$ through $K_1$, $K_2$, $\Omega$, and the Euler parameters, using equations~(\ref{eq:cur_eu1}-\ref{eq:cur_euO}) and the constraint~\eqref{eq:q_constr}, differentiated with respect to $s$.  Third, we eliminate the variables $N_1$ and $N_2$ using equations~\eqref{eq:for_torM} and arrive at the following system of differential equations of 13-th order~\footnote{Here and further, the bars over the variables are dropped for simplicity.}:
\begin{widetext}
\begin{eqnarray}
\label{eq:grand_1}
\alpha \ddot{K_1} &=& (2\beta-1) \dot{(K_2 \Omega)} - \beta K_2 \dot{\Omega} + (\alpha-1) K_1 \Omega^2 + K_1 N_3 + \Omega \dot{g}_2 - \dot{f}_2 - \ddot{g}_1\;,\\
\label{eq:grand_2}
\beta \ddot{K_2} &=& (1-2\alpha) \dot{(K_1 \Omega)} + \alpha K_1 \dot{\Omega} + (\beta-1) K_2 \Omega^2 + K_2 N_3 - \Omega \dot{g}_1 + \dot{f}_1 - \ddot{g}_2\;,\\
\label{eq:grand_3}
\dot{\Omega} &=& (\alpha-\beta) K_1 K_2 - \dot{g}_3 \;,\\
\label{eq:grand_4}
\dot{N}_3 &=& - \alpha \dot{K_1} K_1 - \beta \dot{K_2} K_2 - \dot{\Omega} \Omega  - g_1 K_1 - g_2 K_2 - g_3 \Omega - \dot{f}_3\;,
\end{eqnarray}
\end{widetext}

\begin{widetext}
\begin{eqnarray}
\label{eq:grand_5}
\dot{q_1} &=& \frac{1}{2} ( \phantom{-} K_1 q_4 - K_2 q_3 + \Omega q_2 )\;,\\
\label{eq:grand_6}
\dot{q_2} &=& \frac{1}{2} ( \phantom{-} K_1 q_3 + K_2 q_4 - \Omega q_1 )\;,\\
\label{eq:grand_7}
\dot{q_3} &=& \frac{1}{2} ( - K_1 q_2 + K_2 q_1 + \Omega q_4 )\;,\\
\label{eq:grand_8}
\dot{q_4} &=& \frac{1}{2} ( - K_1 q_1 - K_2 q_2 - \Omega q_3 )\;,\\
\label{eq:grand_9}
\dot{x}   &=& 2 ( q_1 q_3 + q_2 q_4 )\;,\\
\label{eq:grand_10}
\dot{y}   &=& 2 ( q_2 q_3 - q_1 q_4 )\;,\\
\label{eq:grand_11}
\dot{z}   &=& - q_1^2 - q_2^2 + q_3^2 + q_4^2 \;.
\end{eqnarray}
\end{widetext}

The solutions to this system correspond to the equilibrium conformations of the elastic rod. The 13 unknown functions~-- $\rv(s)$, $q_{i=1-4}(s)$, $\vec{N}(s)$, and $\vec{M}(s)$~\footnote{The functions $\vec{M}(s)$ are directly obtainable from $\vec{k}(s)$ by virtue of~\eqref{eq:ber_eul}.}~-- describe the geometry of the elastic rod and the distribution of the stress and torques along the rod.  The equations can be solved for various combinations of initial and boundary conditions.  The case considered in this paper will be the boundary value problem, when the equilibrium solutions are sought for the elastic rod with fixed ends -- that is, with known locations $\rv(0)$, $\rv(1)$ of its ends at $s=0$ and $s=1$, and known orientation $q_{i=1-4}(0)$, $q_{i=1-4}(1)$ of the cross section at those ends. Such case would correspond, for example, to a DNA loop whose ends are bound to a protein.

\begin{figure}[b]
\begin{center}
\includegraphics[scale=.55]{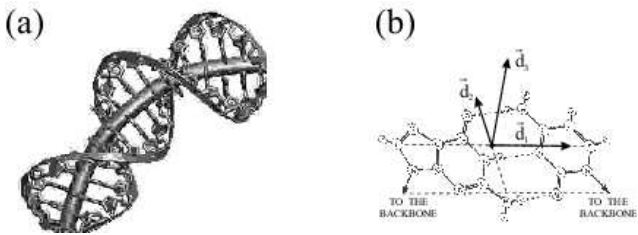}
\end{center}
\caption{\label{fig:elrod_dna}(a) The elastic rod fitted into an all-atom structure of DNA. (b) A coordinate frame associated with a DNA base pair, according to~\cite{OLSO2001}.}
\end{figure}

In general, the system~(\ref{eq:grand_1}-\ref{eq:grand_11}) has multiple solutions for a given set of boundary conditions.  The dimensionless elastic energy of each solution is computed -- according to the Bernoulli-Euler approximation~\eqref{eq:ber_eul} -- as the quadratic functional of the curvatures and the twist:
\begin{equation}
\label{eq:ener_funct}
U = \int_0^1 \left(\frac{\alpha K_1^2}{2} + \frac{\beta K_2^2}{2} + \frac{\Omega^2}{2}\right)ds\;.
\end{equation}
\noindent The straight elastic rod becomes the zero-energy ground state for the functional~(\ref{eq:ener_funct}).  If the interactions of the elastic rod with external bodies, expressed through the forces $\vec{f}$ and torques $\vec{g}$, are not negligible then the energy functional will include additional terms due to those interactions.


\subsection{The elastic rod model for DNA}

Elastic rod theory is a natural choice for building a model of DNA -- a long linear polymer.  The centerline of the rod follows the axis of the DNA helix and Watson-Crick base pairs form cross-sections of the DNA ``rod'' (Fig.~\ref{fig:elrod_dna}\,a). A coordinate frame can be associated with each base pair according to a general convention~\cite{OLSO2001} (Fig.~\ref{fig:elrod_dna}\,b). For a known all-atom structure of a DNA loop, an elastic rod (ribbon) can be fitted into the loop using those coordinate frames. Conversely, a known elastic rod model of the DNA loop can be used directly to build an all-atom structure of the loop (see App.~\ref{sec:appendixf}). Finally, if the all-atom structure is known only for the base pairs at the ends of the loop (as in the case of the \lr\ -- cf Sec.~\ref{sec:lac_solutions}) then the coordinate frames \dframe\ can be associated with those base pairs and provide boundary conditions for Eqs.~\refgrand.  

However, several modifications of the classical theory are necessary in order to describe certain essential properties of DNA.

First of all, the relaxed shape of DNA is a helix, which is described by a tightly wound ribbon rather than a straight untwisted rod treated by our equations so far.  The helix has an average twist of $10^\circ$/\AA\ so that one helical turn takes about 36~\AA.  This is much smaller than the persistent length of DNA bending (500~\AA) or twisting (750~\AA)~\cite{HAGE88,STRI96}, so even a relatively straight segment of DNA is highly twisted.  We introduce the intrinsic twist $\omega^\circ(s)$ of the elastic rod as a parameter in our model.  It is considered to be a function of arclength $s$, because the twist of real DNA varies between different sequences~\cite{OLSO98}.

Second, certain DNA sequences are also known to form intrinsically curved rather than straight helices~\cite{HAGE90}. In terms of our theory that means that the curvature of the relaxed rod may be different from zero in certain sections of the rod.  The intrinsic curvatures $\kiot(s)$ are introduced in our model similarly to the intrinsic twist -- as functions of arclength $s$ determined by the sequence of the DNA piece in consideration.

The intrinsic twist and curvatures result in the modified Bernoulli-Euler equation~\eqref{eq:ber_eul}:
\begin{equation}
\label{eq:ber_eul_1}
\vec{M}(s) = A_1 \kappa_1 \dv + A_2 \kappa_2 \dt + C \omega \dth \;,
\end{equation}
where
\begin{equation}
\label{eq:cur-tw_devs}
\kot(s) = K_i(s) - \kappa_i^\circ(s), \;\; \omega(s) = \Omega(s) - \omega^\circ(s).
\end{equation}

Now, the elastic torques are proportional to the changes in the curvatures and twist from their intrinsic values, rather than to their total values. Note, however, that the ``geometrical'' eqs.~\eqref{eq:fr_rot} -- and consequently, eqs.~(\ref{eq:grand_5}-\ref{eq:grand_8}) -- still contain the full values of the twist and curvature, so that switching from eq.~\eqref{eq:ber_eul} to eq.~\eqref{eq:ber_eul_1} does not simply result in replacing $K_i$ and $\Omega$ with $\kappa_i$ and $\omega$ throughout the system~(\ref{eq:grand_1}-\ref{eq:grand_11}).

Another important structural property of real DNA, which we want to encapsulate in our theory, is the sequence-dependence of the DNA flexibility, i.e., that certain sequences of DNA are more rigid than others~\cite{OLSO96,OLSO2000,MATS2002}.  Accordingly, the bending and twisting rigidities $A_1$, $A_2$, and $C$ in eq.~\eqref{eq:ber_eul_1} become functions of arclength $s$ rather than constants.  The exact shape of these functions depends on the sequence of the studied DNA piece (cf App.~\ref{sec:appendixd}).  The dimensionless bending rigidities $\alpha$, $\beta$, and $\gamma$ (as well as the dimensionless forces and torques) now become scaled not by $C$, but by an arbitrary chosen value of the twisting modulus $\Co$ (for example, the average twisting rigidity of DNA):
\begin{gather}
\label{eq:mod_scale_1}
\alpha = A_1/\Co, \;\; \beta = A_2/\Co, \;\; \gamma = C/\Co \:, \\
\label{eq:for_scale_1}
\bar{N}_{i=1-3} = N_i l^2/\Co, \;\; \bar{M}_{i=1-3} = M_i l/\Co
\end{gather}
\noindent ({\it cf} eq.~(\ref{eq:mod_scale_0},\ref{eq:for_scale_0})).

With the above changes, the new 'grand' system of differential equations becomes: 
\begin{widetext}
\begin{eqnarray}
\label{eq:adv_grand_1}
\ddot{(\alpha \kappa_1)} &=& \phantom{-} \dot{(2 \beta \kappa_2 \Omega)} - \dot{(\gamma K_2 \omega)} - \beta \kappa_2 \dot{\Omega} + \alpha \kappa_1 \Omega^2 - \gamma K_1 \omega \Omega + K_1 N_3 + \Omega \dot{g}_2 - \dot{f}_2 - \ddot{g}_1\;,\\
\label{eq:adv_grand_2}
\ddot{(\beta \kappa_2)} &=& - \dot{(2 \alpha \kappa_1 \Omega)} + \dot{(\gamma K_1 \omega)} + \alpha \kappa_1 \dot{\Omega} + \beta \kappa_2 \Omega^2 - \gamma K_2 \omega \Omega + K_2 N_3 - \Omega \dot{g}_1 + \dot{f}_1 - \ddot{g}_2\;,\\
\label{eq:adv_grand_3}
\dot{(\gamma \omega)} &=& \phantom{-} \alpha \kappa_1 K_2 - \beta K_1 \kappa_2 - \dot{g}_3 \;,\\
\label{eq:adv_grand_4}
\dot{N}_3 &=& - \dot{(\alpha \kappa_1)} K_1 - \dot{(\beta \kappa_2)} K_2 - \dot{(\gamma \omega)} \Omega  - g_1 K_1 - g_2 K_2 - g_3 \Omega - \dot{f}_3\;,\\
\label{eq:adv_grand_5}
\dot{q_1} &=& \frac{1}{2} ( \phantom{-} K_1 q_4 - K_2 q_3 + \Omega q_2 )\;,\\
\label{eq:adv_grand_6}
\dot{q_2} &=& \frac{1}{2} ( \phantom{-} K_1 q_3 + K_2 q_4 - \Omega q_1 )\;,\\
\label{eq:adv_grand_7}
\dot{q_3} &=& \frac{1}{2} ( - K_1 q_2 + K_2 q_1 + \Omega q_4 )\;,\\
\label{eq:adv_grand_8}
\dot{q_4} &=& \frac{1}{2} ( - K_1 q_1 - K_2 q_2 - \Omega q_3 )\;,\\
\label{eq:adv_grand_9}
\dot{x}   &=& 2 ( q_1 q_3 + q_2 q_4 )\;,\\
\label{eq:adv_grand_10}
\dot{y}   &=& 2 ( q_2 q_3 - q_1 q_4 )\;,\\
\label{eq:adv_grand_11}
\dot{z}   &=& - q_1^2 - q_2^2 + q_3^2 + q_4^2 \;,
\end{eqnarray}
\end{widetext}
\noindent and the new energy of the elastic rod is computed as:
\begin{equation}
\label{eq:adv_ener_funct}
U = \int_0^1 \left(\frac{\alpha \kappa_1^2}{2} + \frac{\beta \kappa_2^2}{2} + \frac{\gamma \omega^2}{2}\right)ds\;.
\end{equation}

The system~(\ref{eq:adv_grand_1}-\ref{eq:adv_grand_11}) describes the elastic rod model of DNA in the most general terms.  Not all of the options provided by such model will be explored in the present work; most times the equations will be simplified in one way or another.  The unexplored possibilities and situations when they might become essential will be discussed in Sec.~\ref{sec:discussion}.

To conclude this section, let us observe two immediate results of switching from the classical equations~(\ref{eq:grand_1}-\ref{eq:grand_11}) to the more realistic equations~(\ref{eq:adv_grand_1}-\ref{eq:adv_grand_11}). First, the high intrinsic twist $\omega^\circ(s)$ results in strongly oscillatory behavior of solutions to the system~(\ref{eq:adv_grand_1}-\ref{eq:adv_grand_11}). Due to the non-linear character of the system, the oscillatory component can not be separated from the rest of the solution.  Second, a DNA loop that contains intrinsically bent segments (those inside which $\kappa_{1,2}^\circ \neq 0$) may not be uniformly twisted -- as it would be in the classic case -- even if the loop were isotropically flexible ($\alpha = \beta$) and no external torques were acting upon the loop ($\vec{g}=0$).  Whereas the classical theory necessitates that in such case $\dot{\Omega}=0$ ({\it cf}~eq.~\eqref{eq:grand_3}), the right-hand part of the updated eq.~\eqref{eq:adv_grand_3} is non-zero if $K_{1,2} \neq \kappa_{1,2}$.

 
\section{Elastic Rod Solutions for the DNA Loop clamped by the {\it Lac} Repressor}
\label{sec:lac_solutions}
 
In this section we first describe the test case system for our theory: the complex of the {\it lac} repressor protein with DNA.  Then this protein-DNA system is used to illustrate the numeric algorithm of solving the equations of elasticity. Finally, different solutions for the DNA loop clamped by the {\it lac} repressor are presented.

\begin{figure*}[p]
\begin{center}
\includegraphics[scale=1.]{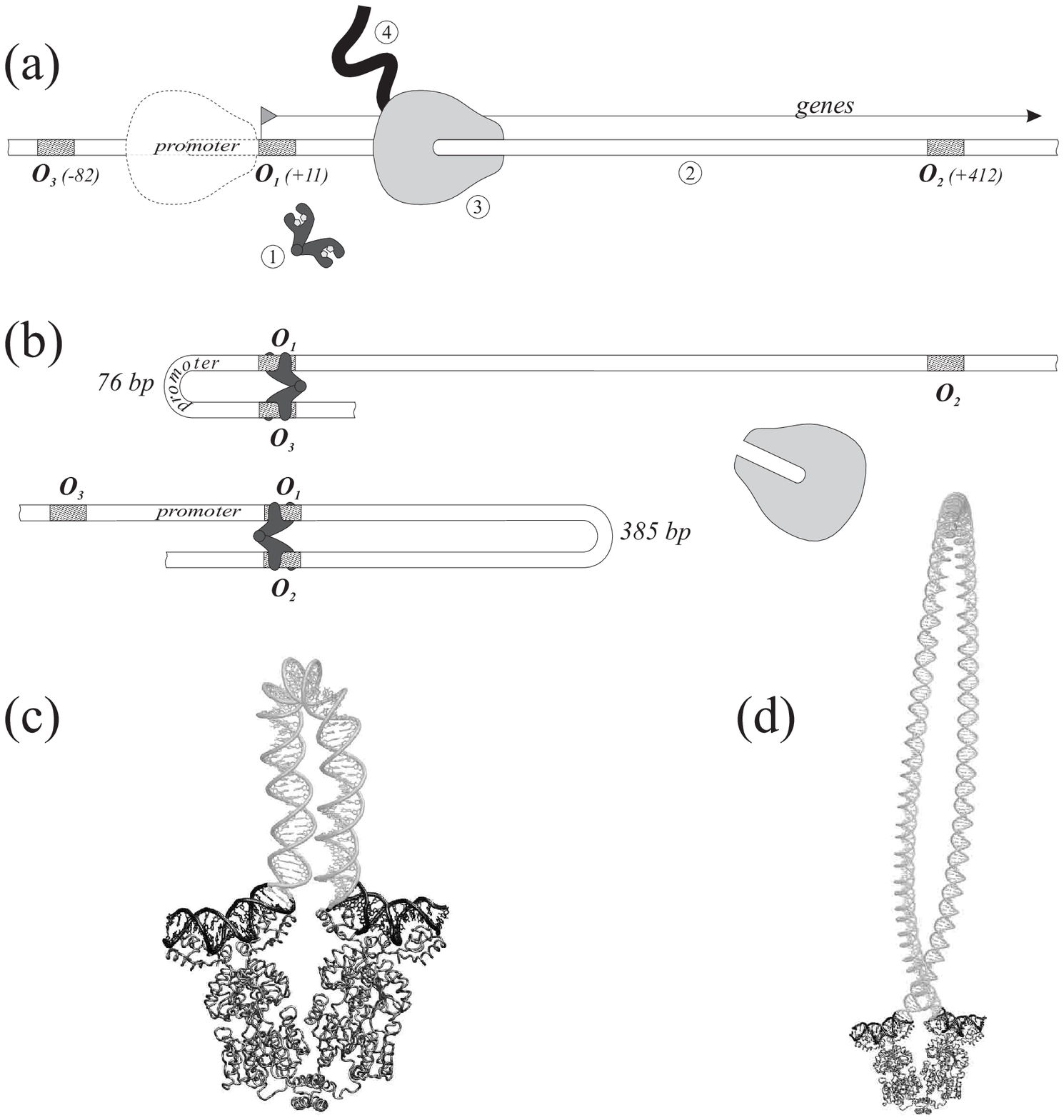}
\end{center}
\caption{\label{fig:lac_operon}(a) The expressed \lo. The biomolecules involved are: (1) - \lr\ (shown deactivated by 4 bound lactose molecules), (2) - the DNA of \ecoli, (3) - RNA polymerase (shown first bound to the promoter and then transcribing the \lo\ genes), (4) - mRNA (shown as being transcribed by the RNA polymerase). The flag shows the ``+1'' base pair of the DNA where the genes of the \lo\ begin. The operator sites, marked as O$_{1-3}$, are shown as shaded rectangles; the position of each operator's central base pair is shown in brackets.  (b) The repressed \lo. The \lr\ is shown binding two operator sites -- either O$_1$ and O$_3$ or O$_1$ and O$_2$ -- and the RNA polymerase -- released from the operon. The end-to-end length of the DNA loop formed in each case is indicated. (c) The crystal structure of the \lr~\cite{LEWI96}. The DNA loop, missing in the crystal structure and shown here in light color, corresponds to an all-atom model fitted to one of the two elastic rod structures predicted for the 76~bp loop (see~\cite{BALA99} and Sec.~\ref{sec:lacsols_short}). Small pieces of the crystal structure are omitted from the figure for clarity. (d) Same as (c) for the 385~bp loop; the all-atom DNA model is fitted to one of the four possible structures of the loop (see Sec.~\ref{sec:lacsols_long} and Fig.~\ref{fig:long_sols}).}
\end{figure*}


\subsection{The DNA complex with the {\it lac} repressor}
\label{sec:lac_operon}

For a test case study, the developed theory was used to build a model of the DNA loop induced in the \ecoli\ genome by the \lacrep\ protein. \Lacrep\ functions as a switch that shuts down the lactose ({\it lac}) operon -- a famous set of \ecoli\ genes, the studies of which laid one of the cornerstones of modern molecular biology~\cite{MILL80,PTAS92,BERG2002}. The genes code for proteins that are responsible for lactose digestion by the bacterium; they are shut down by the \lacrep\ when lactose is not present in the environment. When lactose is present, a molecule of it binds inside the \lr\ and deactivates the protein, thereby inducing the expression of the \lo\ (Fig.~\ref{fig:lac_operon}\,a-b).

The \lr\ consists of two DNA-binding ``hands'', as it can be seen in the crystal structure of the protein~\cite{LEWI96} (Fig.~\ref{fig:lac_operon}\,c).  Each ``hand'' recognizes a specific 21-bp long sequence of DNA, called the operator site.  The \lacrep\ binds to two operator sequences and causes the DNA connecting those sequences to fold into a loop. There are three operator sequences in the \ecoli\ genome: O$_1$, O$_2$, and O$_3$~\cite{OEHL90}; the repressor binds to O$_1$ and either O$_2$ or O$_3$, so that the resulting DNA loop can have two possible lengths: 385~bp (O$_1$-O$_2$) or 76~bp (O$_1$-O$_3$) (Fig.~\ref{fig:lac_operon}\,b).  All three operator sites are necessary for the maximum repression of the \lo~\cite{OEHL90}. While the long O$_1$-O$_2$ loop (385~bp) is the easier to form, the short O$_1$-O$_3$ loop (76~bp) contains the \lo\ promoter~\footnote{Promoter is a regulatory DNA region located upstream from the set of genes it controls and containing binding sites for RNA polymerase and regulatory proteins.} so that folding this 76~bp region into a loop is certain to disrupt the expression of the \lo.

It would hardly be possible to crystallize the DNA loops induced by the \lr\ -- merely because of their size -- and thus the crystal structure~\cite{LEWI96} of the \lr\ was obtained with only two disjoint operator DNA segments bound to the protein~\footnote{The DNA segments were co-crystallized together with the protein for two reasons: (i) to reveal the structure of the DNA-binding domains of the \lr, which are partially unfolded when not binding a piece of DNA~\cite{LEWI96}; (ii) to reveal the structure of the protein-DNA interface.}. The equations of elasticity, discussed above, can be used to build elastic rod structures of the missing loops, connecting the two DNA segments. Such structure would allow to further the study of the \lr-DNA interactions in several ways. First, the force of the protein-DNA interactions computed after solving the equations of elasticity can be used in modeling the changes in the structure of the \lr\ that likely occur under the stress of the bent DNA loop. Second, the elastic rod structure of the loop may serve as a scaffold on which to build all-atom structures of its parts of interest -- such as the binding sites of other proteins, important for the \lo\ expression, e.g., RNA polymerase and CAP -- or, indeed, of the whole loop (cf App.~\ref{sec:appendixf}). All-atom simulations of these sites, either alone or with the proteins docked, may provide interesting keys to the interactions of the regulatory proteins with the bent DNA loop and therefore, to the mechanism of the \lo\ repression. Third, one can predict how changing the DNA sequence in the loop would influence the interactions between the \lr\ and the DNA -- by changing the sequence-dependent DNA flexibility in our model and observing the resulting changes in the structure and energy of the DNA loop.  Finally, the \lr\ system can be used to tune the elastic model of DNA per se, in terms of parameters and complexity level, by comparing the predictions resulting from our model with the experimental data.

These questions will be further discussed in Sec.~\ref{sec:discussion}, while the following sections will detail our study of the \lr\ system.


\subsection{Solving equations of elasticity for the 76~bp-long promoter loop O$_1$-O$_3$}
\label{sec:lacsols_short}

The crystal structure~\cite{LEWI96} of the \lr-DNA complex provides the boundary conditions for the equations of elasticity~(\ref{eq:adv_grand_1}-\ref{eq:adv_grand_11}). The terminal~\footnote{In fact, the boundaries of the loop were placed on the third base pair from the end of each DNA segment. The two terminal base pairs were disregarded, because their structure was seriously disrupted, apparently, due to interactions with solvent. Moreover, the two terminal base pairs do not contact the \lr, so their structure and orientation could easily change within the continuous DNA loop.} base pairs of the protein-bound DNA segments are interpreted as the cross-sections of the loop at the beginning and at the end, and orthogonal frames are fitted to those base pairs, as illustrated in Fig.~\ref{fig:elrod_dna}\,b. The positions of the centers of those frames and their orientations relative to the lab coordinate system (LCS) provide 14 boundary conditions: $\rv(0)$, $\rv(1)$, $\qi(0)$, $\qi(1)$ for equations~(\ref{eq:adv_grand_1}-\ref{eq:adv_grand_11}). In order to match the 13th order of the system, a boundary condition for one of the $q_i$'s is dropped; it will be automatically satisfied because the identity~\eqref{eq:q_constr} is included into the equations.

\begin{figure}[!t]
\begin{center}
\includegraphics[scale=.44]{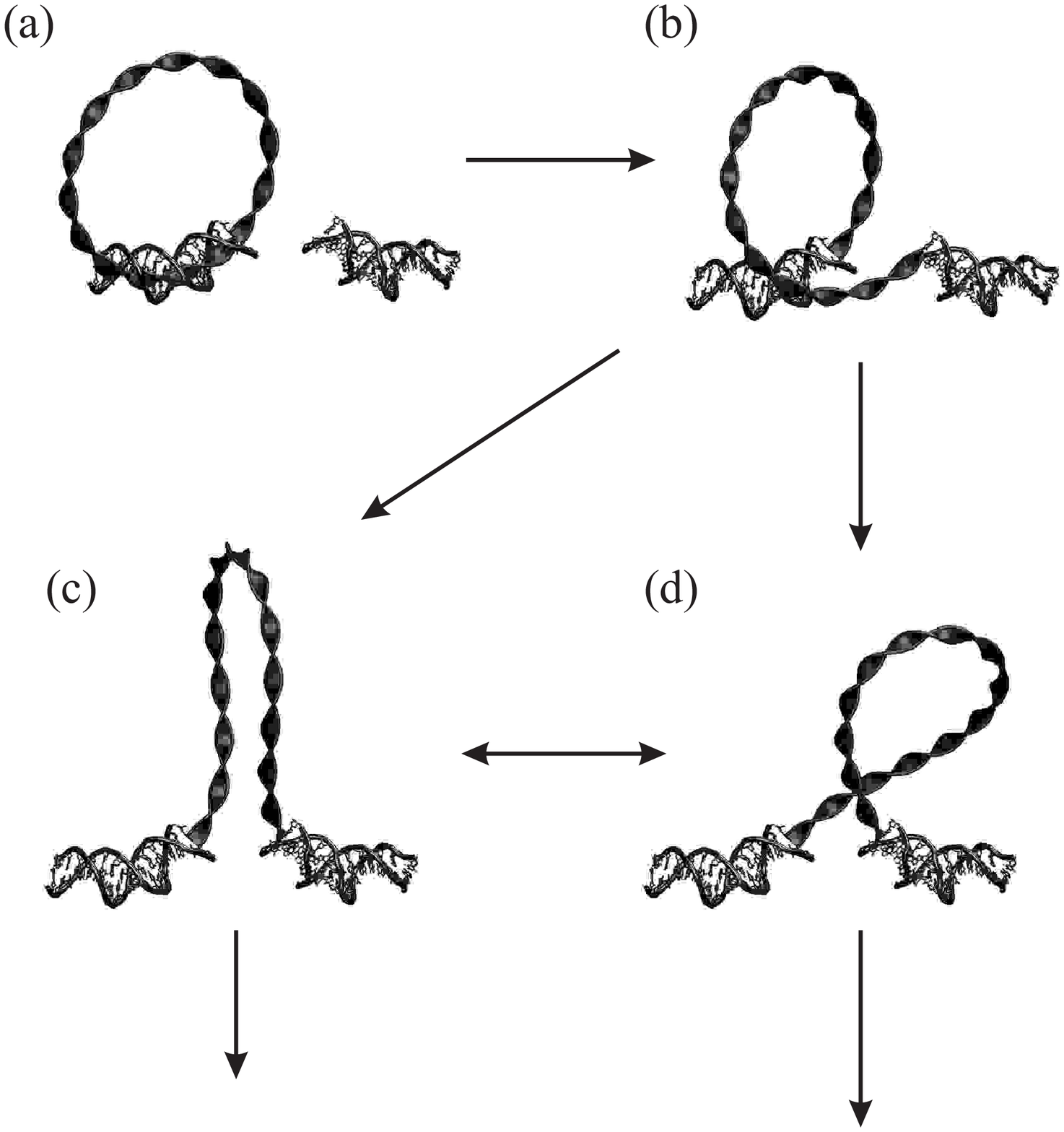}
\includegraphics[scale=.44]{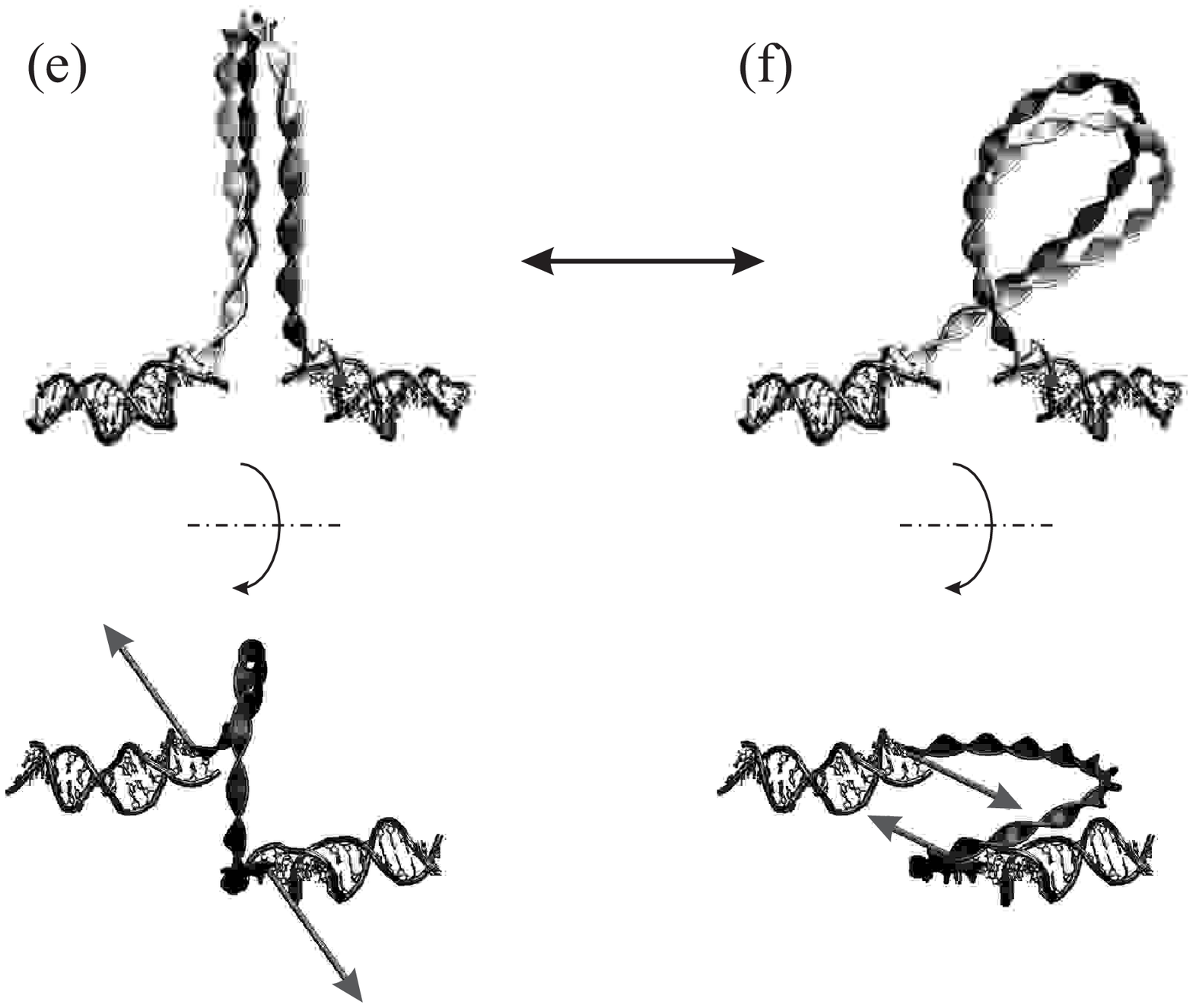}
\end{center}
\caption{\label{fig:iter_short}Evolution of the elastic rod structure during the solution of the BVP for the short loop. (a)~The initial solution: a closed circular loop. (b)~The solution after the first iteration cycle. (c,d)~The solutions after the second iteration cycle, for the clockwise (c) and counterclockwise (d) rotation of the $s=1$ end. (e,f)~The solutions after the third iteration cycle; the previous solutions are shown in light color; the views from the top include the forces that the DNA loop exerts on the DNA segments bound to the \lr. The protein-bound DNA segments from the \lr\ crystal structure are shown for reference only, as they played no role during the iteration cycles except for providing the boundary conditions.}
\end{figure}

The iterative continuation algorithm used for solving the BVP is the same as that used in our previous work~\cite{BALA99} (with some modification when the electrostatic self-repulsion is included into the equations, as described in Sec.~\ref{sec:electrostatics}).  The solution to the problem is constructed in a series of iteration cycles.  The cycles start with a set of parameters and boundary conditions for which an exact solution is known rather than the desired ones. Gradually, the parameters and the boundary conditions are changed towards the desired ones. Usually, only some of the parameters are being changed during each specific iteration cycle: for example, only $\rv(1)$ or only $\alpha/\beta$.  During the cycle, the parameters evolve towards the desired values in a number of iteration steps; the number of steps is chosen depending on the sensitivity of the problem to the modified parameters.  At each step, the solution found on the previous step is used as an initial guess; with a proper choice of the iteration step, the two consecutive solutions are close to each other, which guarantees the convergence of the numerical BVP solver. For the latter, a classical software, COLNEW~\cite{BADE87}, is employed.  COLNEW uses a damped quasi-Newton method to construct the solution to the problem as a set of collocating splines.

The exact solution, from which the iteration cycles started, was chosen to be a circular closed ($\rv(0)=\rv(1)$) elastic loop with zero intrinsic curvature $\kiot$, constant intrinsic twist $\omo$ = 34.6 deg/bp (the average value for the classical B-form DNA~\cite{BERG2002}), constant elastic moduli $\alpha=\beta=\frac{1}{2}$, and no electrostatic charge ($\QD=0$)~\footnote{In principle, the exact solution to begin the iterations with could be obtained by solving Eqs.~\refgrandA\ analytically, as described in~\cite{COLE95}, in the case of an isotropically flexible ($\alpha=\beta$) elastic loop. That would save us the first three iteration cycles. However, for that to be possible, the boundary conditions had to be symmetric, that is, the angles between the normal $\dth$ to the cross-section and the end-to-end vector $\rv(1)-\rv(0)$ had to be the same at both $s=0$ and $s=1$ ends. In our case, the angles in question were equal to 65$\dg$ and 99$\dg$ at the $s=0$ and $s=1$ ends, respectively.}. This solution is shown in Fig.~\ref{fig:iter_short}\,a; the explicit form of the solution is given in App.~\ref{sec:appendixa}.  The loop started (and ended) at the center of the terminal base pair of one of the protein-bound DNA segments.  The coordinate frame associated with that base pair (or the loop cross-section at $s=0$) was chosen as the LCS.  The orientation of the plane of the loop was determined by a single parameter $\psio$ -- the angle between the plane of the loop and the $x$-axis of the LCS.

In the first iteration cycle, the value of $\rv(1)$ was changed, so that the $s=1$ end of the loop moved by 45~\AA\ to its presumed location at the beginning of the second DNA segment (Fig.~\ref{fig:iter_short}\,b). 

In the second iteration cycle, the cross-section of the elastic rod at the $s=1$ end was rotated so as to satisfy the boundary conditions for $\qi(1)$ (Fig.~\ref{fig:iter_short}\,c,d). The rotation consisted in simultaneously turning the normal $\dth$ of the cross-section to coincide with the normal to the terminal base pair and rotating the cross-section around the normal in order to align the vectors $\dv$ and $\dt$ with the axes of the base pair.

\begin{figure}[bt]
\begin{center}
\includegraphics[scale=.4]{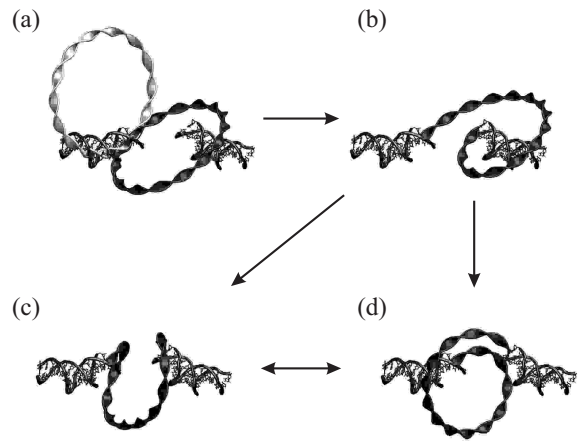}
\end{center}
\caption{\label{fig:iter_concom}Extraneous solutions to the BVP obtained for a different orientation $\psio$ of the initial circular loop. The initial loop from Fig.~\ref{fig:iter_short}\,a is shown in panel (a) in light color.}
\end{figure}

\begin{figure*}[!t]
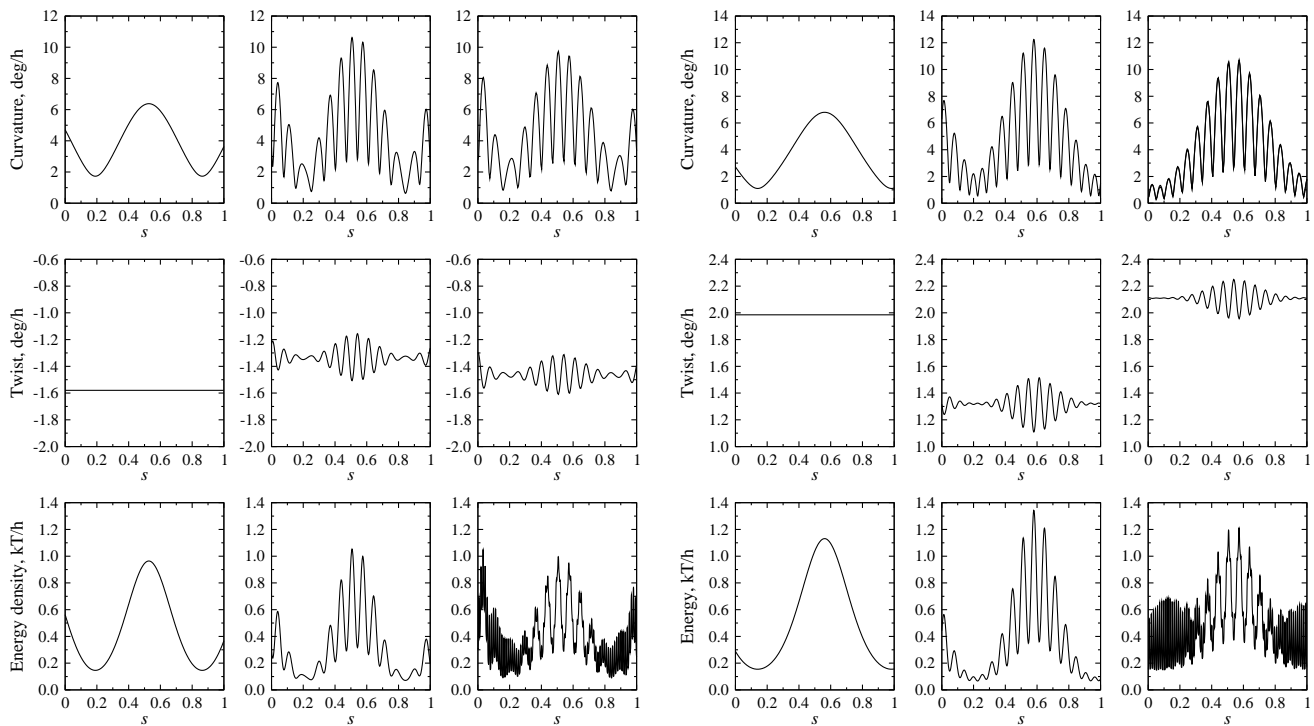

\begin{center}
\includegraphics[scale=0.44]{Fig_6_plot_1}\hspace{5mm}\includegraphics[scale=0.44]{Fig_6_plot_0}
\end{center}
\caption{\label{fig:short_plots}Distribution of curvature, twist, and energy in the elastic rod BVP solutions (U - left panel, O - right panel) for the 76~bp DNA loop. In each panel, left column: the isotropic solution with $\alpha=\beta=2/3$;\: center column: the anisotropic solution with $\alpha=4/15$, $\beta=16/15$;\: right column: the anisotropic solution with electrostatic interactions, for an ionic strength of 10\,mM.}
\end{figure*}

Depending on the direction of the rotation, two different solutions to the problem arise. Rotating the $s=1$ end clockwise results in the solution shown in Fig.~\ref{fig:iter_short}\,c. Rotating the end counterclockwise results in the solution shown in Fig.~\ref{fig:iter_short}\,d. The former solution is underwound by $\omega = -1.4\,\dg$ per bp on the average and the latter solution is overwound by $\omega = 1.6\,\dg$ per bp on the average -- hence, they will be hereafter named ``U'' and ``O''. The two solutions may be transformed into each other during an additional iteration cycle: namely, a rotation of the $s=1$ end clockwise by $2\,\pi$ around its normal transforms the U solution into the O solution, and vice versa. Then, a rotation of the $s=1$ end by another turn restores the U solution, so that a continuous rotation of the $s=1$ end results in switching between the two solutions. That happens because of a self-crossing of the DNA loop, which is not yet prevented in the model at this point.  Topologically, rotating the end increases the linking number~\cite{VOLO94,OLSO99} of the loop by $2\,\pi$ and a self-crossing reduces the linking by the same amount; therefore, two full turns are exactly compensated by one self-crossing, and the original solution gets restored after two turns.

In the third iteration cycle, the bending moduli $\alpha$ and $\beta$ (so far, equal to each other) were changed from 0.5 to 2/3 -- which is the ratio between DNA bending and twisting moduli that most current experiments agree upon~\footnote{The persistence length of DNA bending has universally been measured to be around $L_A =$ 500~\AA~\cite{HAGE88,BAUM97}.  There is less agreement as to what the twisting persistence length should be, but most of the recent data agree on the value of $L_C =$ 750~\AA~\cite{STRI96}.  From the relations for the bending and twisting moduli $A = L_A k T$ and $C = L_C k T$~\cite{MARK94} one obtains $A = 1.5 \cdot 10^{-19} $erg$\cdot$cm, $C = 3 \cdot 10^{-19} $erg$\cdot$cm, and $A/C = L_A/L_C =$ 2/3.}. Such increase in the bending rigidity slightly changed the geometry of the U and O solutions (Fig.~\ref{fig:iter_short}\,e-f) and increased the unwinding/overwinding to $-1.6\,\dgbp$ and $2.0\,\dgbp$, respectively. The change in $\omega$ has a clear topological implication: the deformation of the looped DNA is distributed between the writhe (bending) of the centerline and the unwinding/overwinding of the DNA helix.  When the bending becomes energetically more costly, the centerline of the loop straightens up (on the average) and the deformation shifts towards more change in twist.

Notably, two more solutions may result from the iteration procedure, depending on the orientation $\psio$ of the initial simplified circular loop (Fig.~\ref{fig:iter_concom}).  However, for the 76~bp loop these solutions are not acceptable, because the centerlines of the corresponding DNA loops would have to run right through the structure of the \lr\ (cf Fig.~\ref{fig:iter_concom}\,c-d and Fig.~\ref{fig:lac_operon}\,c).

Therefore, only the two former solutions to the problem are acceptable in the case of the short loop. The shapes of the solutions obtained after the third iteration cycle become our first-approximation answer to what the structure of the DNA loop created by the \lr\ must be. The solutions are portrayed in Fig.~\ref{fig:iter_short}\,e-f, and the profiles of their curvature, twist, and elastic energy density are shown in Fig.~\ref{fig:short_plots} (left columns of the two panels).

The U solution forms an almost planar loop, its plane being roughly perpendicular to the protein-bound DNA segments (Fig.~\ref{fig:iter_short}\,e). The shape of the loop resembles a semicircle on two relatively straight segments connected by short curved sections to the \lr-bound DNA.  Accordingly, the curvature of the loop is highest in the middle and at the ends, the strongest bend being around 6~deg/bp, and drops in between (Fig.~\ref{fig:short_plots}).  The average curvature of this loop is 3.7~deg/bp.  The unwinding is constant, by virtue of $\alpha = \beta$ -- for the same reason the energy density profile simply follows the curvature plot.  The total energy of the loop is 33.0\,kT, of which 26.8\,kT are accounted for by bending, and 6.2\,kT by twisting.  The stress of the loop pushes the ends of the protein-bound DNA segments (and consequently, the \lr\ headgroups) away from each other with a force of about 10.5\,pN (Fig.~\ref{fig:iter_short}\,e).

The O solution leaves and enters the protein-bound DNA segments in almost straight lines, connected by a semicircular coil of about the same curvature as that of the U solution, not, however, confined to any plane (Fig.~\ref{fig:iter_short}\,f). The average curvature equals 3.6~deg/bp. The energy of this loop is higher: 38.2\,kT, distributed between bending and twisting as 28.5\,kT and 9.7\,kT, respectively.  The forces of the loop interaction with the protein-bound DNA segments equal 9.2\,pN and are pulling the ends of the segments past each other (Fig.~\ref{fig:iter_short}\,f).

Since the energy of the U loop is 5\,kT lower than that of the O loop, one could conclude that this form of the loop should be dominant under conditions of thermodynamic equilibrium.  Yet, both energies are too high: the estimate of the energy of the 76~bp loop from the experimental data~\cite{HSIE87} is approximately 20\,kT at high salt concentration (see App.~\ref{sec:appendixb}). Therefore, one cannot at this point draw any conclusion as to which loop structure prevails, and further improvements to the model are needed, such as those described in sections~\ref{sec:anisotropy}-\ref{sec:electrostatics}.


\subsection{Solutions for the 385~bp-long O$_1$-O$_2$ loop}

\label{sec:lacsols_long}

\begin{figure}[!tbp]
\begin{center}
\includegraphics[scale=.45]{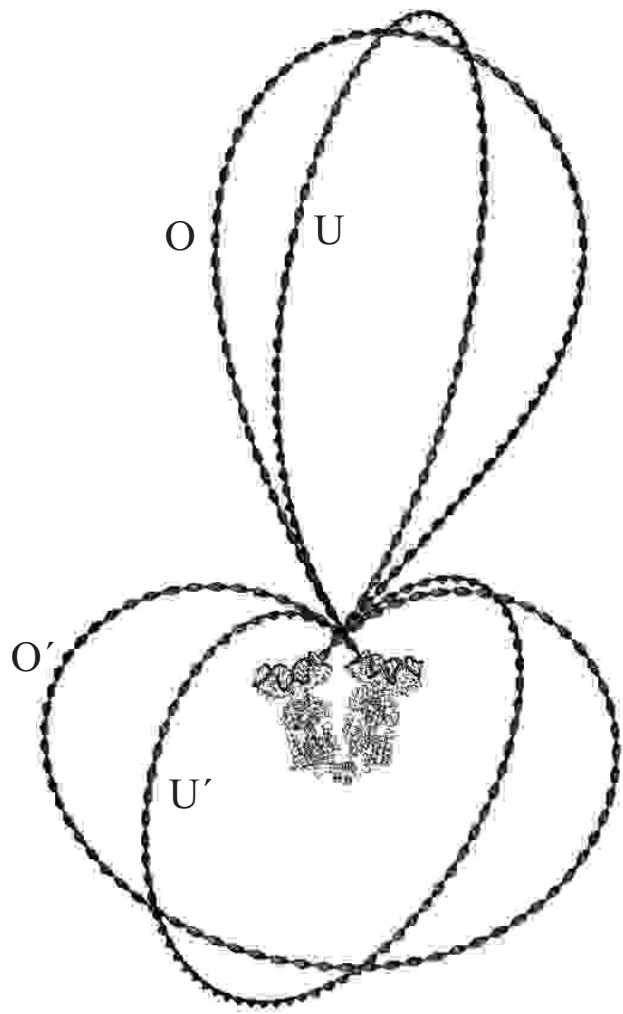}
\end{center}
\caption{\label{fig:long_sols}Four solutions of the BVP problem for the 385~bp-long DNA loop. Underwound solutions are marked by the letter `U', overwound ones, by `O'.}
\end{figure}

Using the same algorithm, the BVP was solved for the 385~bp loop. Similarly, four solutions were obtained (cf Figs.~\ref{fig:iter_short}\,e-f, \ref{fig:iter_concom}\,c-d). With the longer loop, the previously inacceptable solutions are running around the \lr\ rather than through it -- and therefore, are acceptable. All four solutions (denoted as U, \UP, O, \OP) are shown in Fig.~\ref{fig:long_sols}.  The solutions U and \UP\ are underwound, O and \OP\ are overwound.  The geometric and energetic parameters of the four solutions are shown in Table~\ref{tab:long_sols}.

\begin{table}[b]
\begin{ruledtabular}
\begin{tabular}{ccccccc}
Solu- & $<\omega>$ & $<\kappa>$ & $\kappa_{max}$ & $U$ & $U_{\kappa}/U$ & $U_{\omega}/U$ \\
tion & (\dbp) & (\dbp) & (\dbp) & (\mbox{\scriptsize kT}) & & \\
\hline
U & -0.24 & 0.73 & 1.31 & 6.2 & 0.88 & 0.12 \\
O & 0.40 & 0.80 & 1.49 & 8.7 & 0.76 & 0.24 \\
\UP & -0.33 & 1.21 & 1.59 & 14.4 & 0.90 & 0.10 \\
\OP & 0.42 & 1.23 & 1.58 & 15.5 & 0.86 & 0.14 \\
\end{tabular}
\end{ruledtabular}
\caption{\label{tab:long_sols}Geometric and energetic properties of the four solutions of the BVP problem for the 385~bp-long DNA loop, in the case $\alpha = \beta = 2/3$.}
\end{table}

It is, of course, not surprising that the elastic energy and the average curvature and twist of the longer loops are smaller than those of the shorter loops of the same topology. The curvature and the twist decrease because the same amount of topological change (linking number) has to be distributed over larger length -- and the energy density decreases as the square of the curvature/twist, so the integral energy is roughly inversely proportional to the length of the loop. More interestingly, it is the U loop again that has the lowest energy and, on the first look, should be predominant under thermodynamic equilibrium conditions. And the formerly extraneous \UP\ and \OP\ solutions clearly have such high energies that they should practically not be represented in the thermodynamical ensemble of the loop structures and could be safely discounted for the 385~bp loop as well.

Yet again, the conclusions, based on the obtained energy values, shall be postponed until the present elastic rod model is further refined.
 
 
\section{Effects of anisotropic flexibility}
\label{sec:anisotropy}

As the first step towards refining our model, a closer look is taken into the DNA bending moduli $A_1$ and $A_2$ (or, $\alpha$ and $\beta$). The majority of the earlier treatments~\cite{MARK94,MARK95A,MARK98,SHI95,WEST97,KATR97,TOBI2000,COLE2000} used the approximation of isotropic flexibility: $\alpha = \beta$, which holds for an elastic rod with a cross-section that has a rotational symmetry of the 4-th order.  Such approximation simplifies the problem, for example, resulting in a uniform distribution of the extra twist (unwinding or overwinding) $\omega$ along the DNA in the absence of the external twisting moment~$g_3$: $\omega(s) = \mbox{\it const}$ ({\it cf}\ eqs.~\eqref{eq:grand_3},~\eqref{eq:adv_grand_3}).  However, the DNA cross-section is rotationally asymmetric (see Fig.~\ref{fig:elrod_dna}\,b).  Therefore the DNA is treated here as an anisotropically flexible rod -- one with $\alpha \neq \beta$.  Having started with the isotropic model, we consider in this Section the effect of anisotropic flexibility on the conformation and energy of the elastic rod.


\subsection{Anisotropic moduli of DNA bending}
\label{sec:anisotropy_theory}

The bending and twisting moduli (or persistence lengths) $A$ and $C$ of DNA has been measured in many experiments~\cite{HAGE88,HEAT96,STRI2000}.  Most experiments currently agree on the ratio $A/C = 2/3$.  However, this value is obtained by interpreting the experimental data in terms of isotropically bendable DNA, i.e., idealized DNA with a circular cross-section.  However, the DNA structure clearly shows that at the atomic level the DNA bending should be anisotropic: for example, bending of DNA towards one of its grooves should clearly take less energy than bending over its backbone (Fig.~\ref{fig:elrod_dna}).  Here we model the anisotropic rigidity by using two bending moduli $A_1$ and $A_2$ for bending in the direction of the grooves or in the direction of the backbone, respectively. (Other approximations are also conceivable~\cite{OLSO93,OLSO2000}).

How can the experimentally estimated effective bending modulus $A$ be related to the two bending moduli? To obtain a correct answer one should remember that the discussed bending takes place in a tightly twisted rod, i.e., one period of the intrinsic twist of the rod is much smaller than the typical radius of curvature.  In a rod without such intrinsic twist, the thermal fluctuations of bending in two principal directions (that is, of bending around $\dv$ and $\dt$ -- see Fig.~\ref{fig:elrod_dna}\,b) are independent, and a well-known formula (see, e.g., \cite{FLOR69,OLSO93}) is obtained from the principle of energy equidistribution:
\begin{equation}
\frac{1}{A} = \frac{1}{2} \left( \frac{1}{A_1} + \frac{1}{A_2} \right) \;.
\label{eq:bendmod_notw}
\end{equation}

However, when the anisotropic rod is tightly twisted, the thermal fluctuations will inevitably cause bending in both principal directions.  One may consider, for example, the bending a long screw: both the groove and the ridge of its thread will in turn be facing the direction of the bend.  Since unwinding involves energetic penalty as well, the rod can not unwind so as to face the bend only with the ``soft side''. Accordingly, the effective bending modulus $A$ will depend on both $A_1$ and $A_2$.  In the rod with no intrinsic curvature, the first approximation yields the following simple relation:
\begin{equation}
A = \frac{A_1 + A_2}{2}
\label{eq:bendmod_tw}
\end{equation}
\noindent (see App.~\ref{sec:appendixc}).

This single equation is however not sufficient to derive both $\alpha$ and $\beta$ from the $A/C$ ratio as the ratio of $A_1/A_2=\alpha/\beta$ remains unknown.  In~\cite{OLSO93}, the value of $\alpha/\beta = 1/4$ was suggested as both being close to experimental data and reproducing well the DNA persistence length in Monte Carlo simulations. Other estimates result from comparison of the oscillations of roll and tilt -- the angles of DNA bending in the two principal directions~\cite{OLSO98,GORI95} -- which are directly related to $\alpha$ and $\beta$. Moreover, there is always an uncertainty due to the dependence of both the $\alpha/\beta$ and the $A/C$ ratio on the DNA sequence -- and for a specific short DNA loop, such as the one studied here, they may differ from the average values measured for a long DNA with variable sequence.

For these reasons, we decided to study the effect of bending anisotropy on the structure and energy of the \lr\ loops in a broad range of parameters $\mu=\alpha/\beta$ and $A/C \approx (\alpha+\beta)/2$, and to pick the loops with $A/C=2/3$ and $\mu = 1/4$ for a detailed structural and energetic analysis.


\subsection{Changes to the {\it lac} repressor loops due to bending anisotropy in the case of $\mu=1/4$}
\label{sec:anisotropy_structure}

Let us first consider the structure of the U and O loops in the specific case $\mu = 1/4$, for it will be easier then to understand the general picture afterwards.  To obtain the new loops, another iteration cycle was performed, during which the previously generated isotropic structures with $\alpha = \beta = 2/3$ were transformed by simultaneously changing the bending moduli $\alpha$ and $\beta$ towards the desired values of $4/15$ and $16/15$, respectively.

\begin{figure}[!b]
\begin{center}
\includegraphics[scale=.45]{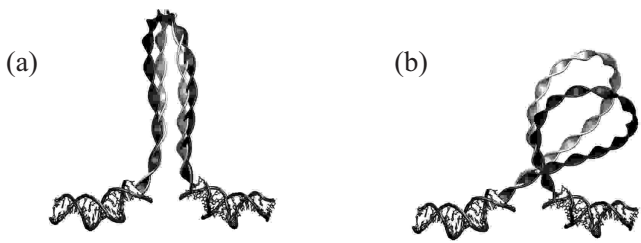}
\end{center}
\caption{\label{fig:aniso_drw}Changes in the predicted structure of the elastic loop due to the bending anisotropy. (a) The underwound solution. (b) The overwound solution. Structures obtained for $\alpha = \beta = 2/3$ are shown in light; structures obtained for $\alpha = 4/15$, $\beta = 16/15$ are shown in dark.}
\end{figure}

The structures of the loops are shown in Fig.~\ref{fig:aniso_drw}. As one can see, the U loop did not change much, the root-mean square deviation (\rmsd)~\footnote{To characterize the difference between two elastic loop solutions $L$ and $L^\prime$ for the \lr\ problem, this paper employs the root-mean square deviation between the centerlines $\rv(s)$ and $\rv^\prime(s)$ of those loops, defined as follows: $\rmsd = \int_0^1 \left| \rv(s) - \rv^\prime(s) \right| ds$.  Typically, the \rmsd\ will be measured in DNA helical steps $h$=3.4\,\AA.} from the isotropic solution being only 1.1\,$h$. In contrast, the O loop changed by 4.6\,$h$ and bent over itself at the point of near self-contact even more than the ``soft'' isotropic loop with $\alpha = \beta = 1/2$ (cf. Fig.~\ref{fig:iter_short}\,f).

The energies of both loops changed significantly, getting reduced by about one-third. The energy of the U loop dropped from 33.0\,kT to 23.3\,kT, distributed between the bending and twisting as 18.9\,kT and 4.4\,kT, respectively. The energy of the O loop dropped from 38.2\,kT to 26.5\,kT, distributed between the bending and twisting as 22.2\,kT and 4.3\,kT, respectively.

To understand the considerable energy change, one should consider the structure of the loops on the small scale.  The distributions of curvature and twist in the loops are shown in Fig.~\ref{fig:short_plots} (central columns).  As one can see, the previously smooth profiles acquired a seesaw pattern.  This happens because the elastic rod -- now better called elastic ribbon -- twists around the centerline with a high frequency due to its high intrinsic twist -- and therefore the vectors $\dv$ and $\dt$ get in turn aligned with the main bending direction (the principal normal $\vec{n}$, see Fig.~\ref{fig:elrod}). In DNA terms, the loop is facing the main bending direction with the grooves and the backbone, in turn (see Fig.~\ref{fig:elrod}).

When the vector $\dt$ is aligned with $\vec{n}$, all the bending occurs towards the grooves, so that $|K_1| = |K|$ and $K_2 = 0$, and the local energy density equals $dU_{\kappa} = \alpha K_1^2/2$.  Then, after a half-period of the intrinsic twist, $\dv$ gets aligned with $\vec{n}$ and all the bending occurs towards the backbone so that $K_1 = 0$, $|K_2| = |K|$, and $dU_{\kappa} = \beta K_2^2/2$. Since $\alpha < \beta$, the latter orientation results in higher energetic penalty and unbending moments than the former one. Therefore, the sections of the rod facing the main bending direction with $\dv$ straighten up and those facing it with $\dt$ become more strongly bent.  The resulting oscillations of the curvature are seen in Fig.~\ref{fig:short_plots}. The structure of the rod becomes an intermediate between that of a smoothly bent loop and of a chain of straight links, the more so the closer the $\alpha/\beta$ ratio gets to 0.  The sections where the rod is facing the main bending direction with $\dt$ play the role of the ``soft joints'' where most of the bending accumulates.

It is this concentration of the curvature in the ``soft joints'' that results in the decrease in the elastic energy of the rod. The energy density oscillates together with the curvature, reaching maxima at the points of the maximum bend (Fig.~\ref{fig:short_plots}).  Since the bending becomes cheaper, some of the rod twist gets redistributed into the bend and the absolute value of the average twist $\omega$ decreased for both U and O solutions (Fig.~\ref{fig:short_plots}), to $-1.3\,\dgbp$ and $1.3\,\dgbp$, respectively.

Locally, the twist $\omega$ of the anisotropically flexible rod also displays oscillations (Fig.~\ref{fig:short_plots}).  When all the bending occurs towards $\dv$, the twist slightly increases, as to wind the ``rigid face'' away from the main bending direction.  When the bending occurs towards $\dt$, the twist slightly decreases as to prolong the exposure of the ``soft face'' to the main bending.  These oscillations of the twist are however not large, because they inflict a certain energetic penalty as well.

Finally, the force of the protein-DNA interaction dropped to 7.9\,pN for the U solution and to 7.2\,pN for the O solution.  The direction of the force in the U solution changed insignificantly; the force in the O solution rotated ``upwards'' by about 40 degrees.


\subsection{Changes to the 76~bp {\it lac} repressor loops in the broad range of parameters $\mu$ and $A/C$}
\label{sec:anisotropy_range_short}

From the described specific case of $\mu = 1/4$ and $A/C=2/3$, we proceeded to studying the elastic rod conformations in a broad range of parameters $A/C$ and $\mu$ (or, $\alpha$ and $\beta$). $A/C$ was varied between 1/20 and 20, and $\mu$ between $10^{-2}$ and $10^2$. In principle, such range is too broad as neither the DNA rigidity $A/C$ significantly deviates from the range of 1, even by alternative estimates~\cite{HAGE88,STRI2000,HEAT96,MATS2002}, nor is $\mu$ likely to change from 1 by two orders of magnitude as the oscillations of DNA roll and tilt are normally of the same order~\cite{GORI95,OLSO98}.  The values of $\mu>1$ are especially unlikely as the DNA bending towards the grooves should clearly take less energy than bending towards the backbone (cf Fig.~\ref{fig:elrod_dna}). Yet, the computations using the developped method were inexpensive, the data could be easily generated, and the broad range of parameters was studied for the sake of generality of our results in regard to the behavior of twisted elastic~rods.

\begin{figure*}[p]
\vskip 1cm
\begin{center}
\includegraphics[scale=.72]{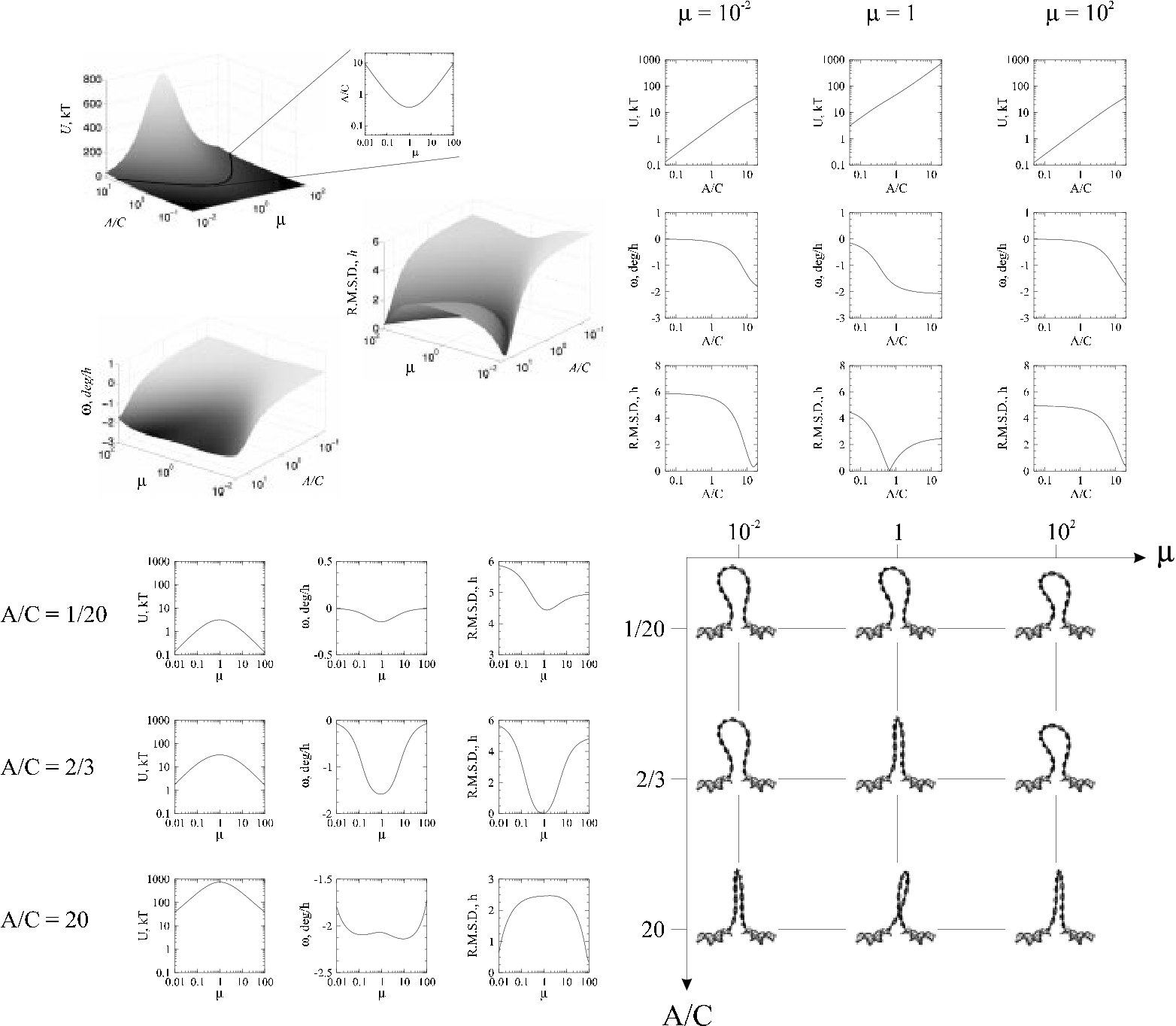}
\end{center}
\vskip 1cm
\caption{\label{fig:range_1}Energy and geometry of the 76~bp U solution in the broad range of elastic moduli $\alpha$ and $\beta$. The plots are made in coordinates $\mu = \alpha/\beta$ and $A/C \approx (\alpha+\beta)/2$~\eqref{eq:bendmod_tw}. Top left quadrant: 3D plots for the elastic energy $U$, the average unwinding $\omega$ of the loop, and the \rmsd\ of the loop centerline from that in the isotropic case $\mu=1$. To present the best view, the orientaion of the plots for \rmsd\ and $\omega$ are shown from a different viewpoint than $U$. The dark contour on the energy map shows the cross-section of the map at 20\,kT, which is the DNA looping energy estimated from experiment~\cite{HSIE87}. The projection of the 20\,kT contour on the $\mu$ -- $A/C$ plane is shown to the right of the energy map. Top right quadrant: cross-sections of the maps for $\mu$ = $10^{-2}$, 1, and $10^2$ (left to right).  Bottom left quadrant: cross-sections of the maps for $A/C$ = 1/20, 2/3, and 20 (top to bottom).  Bottom right quadrant: snapshots of the loop structures arranged according to their $\mu$ values (left to right) and $A/C$ values (top to bottom).}
\end{figure*}

\begin{figure*}[p]
\vskip 1cm
\begin{center}
\includegraphics[scale=0.72]{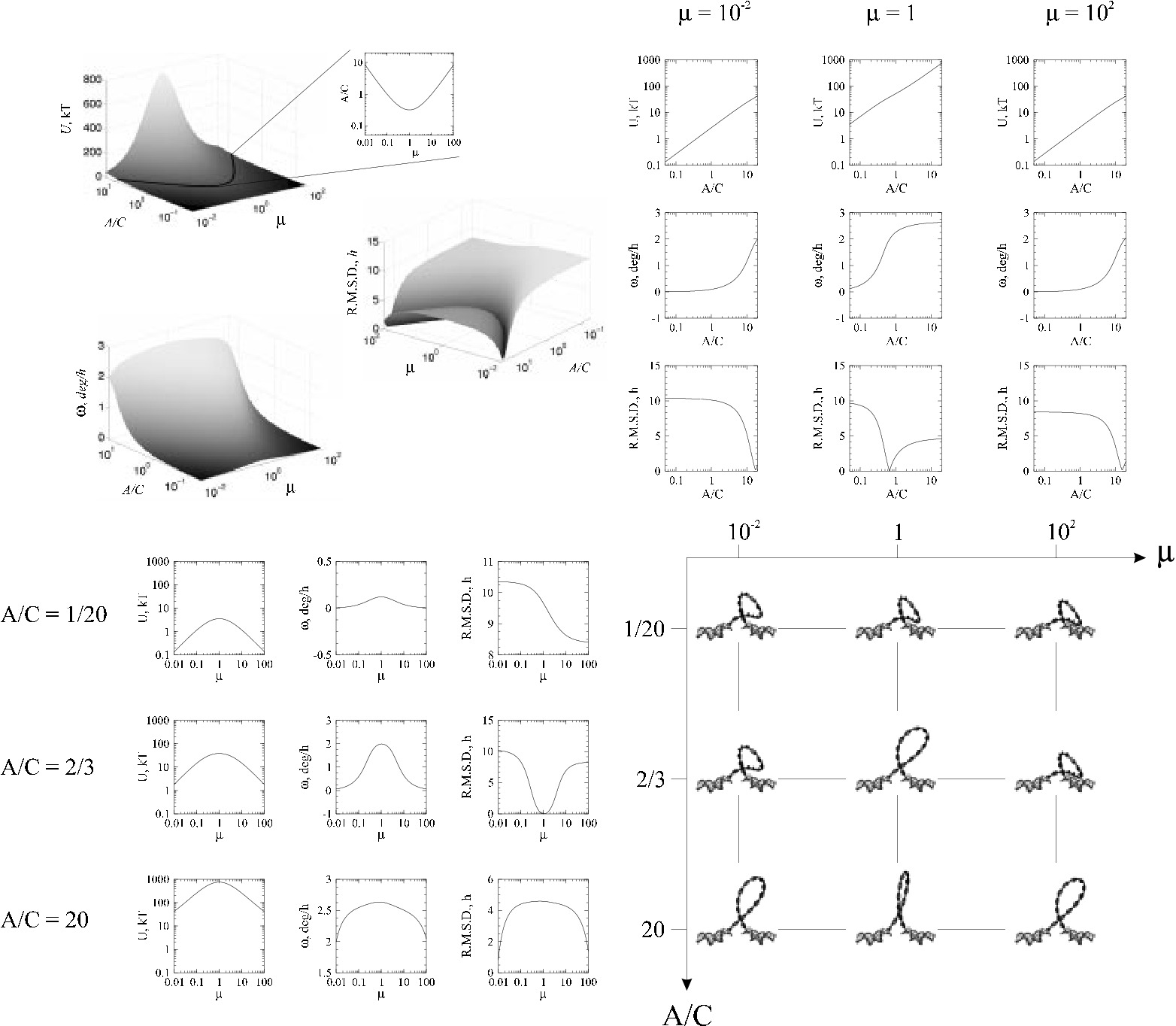}
\end{center}
\vskip 1cm
\caption{\label{fig:range_0}Energy and geometry of the 76~bp O solution in the broad range of elastic moduli $\alpha$ and $\beta$. The plots are made in coordinates $\mu = \alpha/\beta$ and $A/C \approx (\alpha+\beta)/2$~\eqref{eq:bendmod_tw}. Top left quadrant: 3D plots for the elastic energy $U$, the average unwinding $\omega$ of the loop, and the \rmsd\ of the loop centerline from that in the isotropic case $\mu=1$. To present the best view, the orientation of the plot for \rmsd\ is shown from a different viewpoint than $U$ and $\omega$. The dark contour on the energy map shows the cross-section of the map at 20\,kT, which is the DNA looping energy estimated from experiment~\cite{HSIE87}. The projection of the 20\,kT contour on the $\mu$ -- $A/C$ plane is shown to the right of the energy map. Top right quadrant: cross-sections of the maps for $\mu$ = $10^{-2}$, 1, and $10^2$ (left to right).  Bottom left quadrant: cross-sections of the maps for $A/C$ = 1/20, 2/3, and 20 (top to bottom).  Bottom right quadrant: snapshots of the loop structures arranged according to their $\mu$ values (left to right) and $A/C$ values (top to bottom).}
\end{figure*}

The parameters were changed in two ``sweeping'' iteration cycles, which started from the isotropic solutions with $\alpha = \beta = 2/3$ and went all over the studied range. $A/C$ was changed in the first cycle, and $\mu$ -- in the second, nested cycle. $\alpha$ and $\beta$ were computed accordingly on each step, and the new solutions were generated. The results of the computations are presented in Figs.~\ref{fig:range_1}, \ref{fig:range_0} for the U and O solutions, respectively.

Expectedly, the energy $U$ of the elastic rod grows when the bending rigidity $A/C$ is growing. As the relative energetic cost of the bending and twisting changes, the elastic rod changes its shape as to optimally distribute the deformation between bending and twisting. For example, when $A/C$ is growing, the rod becomes less bent and more twisted on the average, the loop straightens itself up and the average unwinding/overwinding $\omega$ increases (Figs.~\ref{fig:range_1}, \ref{fig:range_0}).  Conversely, when $A/C$ goes down, the costly twisting falls to zero and the rod centerline becomes more significantly bent at every point.  Yet, the rod can not straighten itself to a line, nor can $|\omega|$ fall below zero -- therefore, at some point the structure of the rod approaches an asymptotic shape, the average unwinding/overwinding and the \rmsd\ from the initial structure become constant, and the energy becomes simply a linear function of the bending modulus (cf. the plots in Figs.~\ref{fig:range_1}, \ref{fig:range_0} for $\mu=1$). A similar effect has been observed in the studies of the bending anisotropy of a M\"obius band~\cite{MAHA93}.

As in the discussed specific case of $\mu=1/4$, introducing the anisotropic flexibility results in a significant drop of the elastic energy, by as much as an order of magnitude. The bending concentrates in the regions where the loop faces the main bending direction with its ``soft'' side, and the twist develops oscillations.  The average twist normally decreases in its absolute value, as described for $\mu=1/4$, except when the $A/C$ rigidity becomes really large. Then it is energetically costly to increase bending even in the soft spots, and the absolute value of the average twist at first increases (see the $\omega$ plot for $A/C=20$ in Fig.~\ref{fig:range_1}).  Yet, the further increase in the bending anisotropy further reduces $\alpha$ (or $\beta$, if $\mu \rightarrow \infty$) and makes bending in the ``soft spots'' cheaper than twisting, so the absolute average twist eventually turns down after the initial increase. As the twist reduces to zero, the structure of the rod eventually reaches an asymptotic shape, as in the case of the changed bending rigidity.  The changes in structure and energy for $\mu<1$ and $\mu>1$ are nearly symmetric, because in the latter case the ``soft spots'' simply move along the loop centerline by a quarter of a period of the DNA helix -- which is a mere 1/30-th of the total length of the loop.

The cut through the 3D plot for the elastic energy at $U=20\,\kT$ shows that the experimentally estimated looping energy of 20\,kT can be achieved in a wide family of $A/C$ and $\mu$ -- or, $\alpha$ and $\beta$ -- parameters. The parameter families for the U and O solutions are shown as contours on the $\mu$--$A/C$ plane in Figs.~\ref{fig:range_1}, \ref{fig:range_0}. For the standard value of $A/C=2/3$, the experimental energy is achieved for $\mu \approx 1/5$ (or, $\mu \approx 5$). Yet, it is evident that the model of anisotropically flexible twisted DNA can match the bending energy data with different sets of the bending moduli -- so, more data from different experiments are needed in order to conclusively determine what specific moduli should be used.

\begin{figure}[tb]
\begin{center}
\includegraphics[scale=.43]{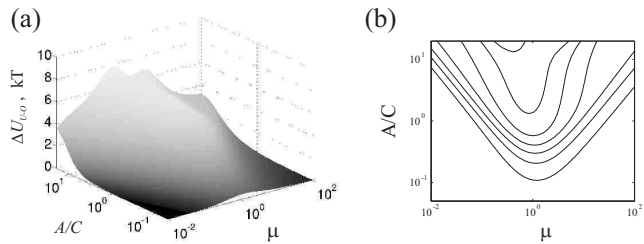}
\end{center}
\caption{\label{fig:uo_diff}(a)~3D plot of the elastic energy difference $\Delta U = U_{\rm U} - U_{\rm O}$ between U and O solutions, in the coordinates $\mu = \alpha/\beta$ and $A/C \approx (\alpha+\beta)/2$~\eqref{eq:bendmod_tw}. (b)~Contours of the plot in (a) for the $\Delta U$ values of $1, 2, \dotsc , 7\,\kT$.}
\end{figure}

It should also be noted here that the approximation $A/C \approx (\alpha+\beta)/2$ was derived with the assumption that the bending of the rod is smooth enough compared to the intrinsic twist. The increase in the bending anisotropy, however, increases the curvature in the soft spots and eventually our assumption breaks down, i.e., the average $(\alpha+\beta)/2$ does not correctly describe the rod rigidity any more for the strongly anisotropic case. This adds uncertainty to the choice of bending moduli $\alpha$ and $\beta$ and necessitates studying the parameters in the broad range.

Yet, certain conclusions about the \lr\ loops can be made even without certainty about the values of $\alpha$ and $\beta$. The difference in the energy between the U and O loops, shown in Fig.~\ref{fig:uo_diff}, exceeds 1\,kT for most of the studied range of $\alpha$ and $\beta$, and in a significant part of the range even goes above 3\,kT. Therefore, it can be safely concluded that, under thermal equilibrium, the DNA loop formed by the \lr\ should preferably have the shape of the U solution.  Incidentally, the shape of this loop changes less with the change in $\alpha$ and $\beta$ (cf Figs.~\ref{fig:range_1}, \ref{fig:range_0}) -- and therefore can with more certainty be used to determine such large-scale parameters of the loop as, for example, the radius of gyration or an average protein-DNA distance.

Interestingly, as the rigidity $A/C$ is increased, the O and U loops converge to the same asymptotic shape (cf snapshots for $A/C=20$, $\mu=1$ in Figs.~\ref{fig:range_1}, \ref{fig:range_0}) -- the shape that has the least possible bending. The only difference is in the twist -- the total twist of the O solution is $2\,\pi$ plus that of the U solution -- and in this case the whole energy difference of about 7\,kT comes from the difference in the twisting energy.

Another notable feature of the O solution is a near self-contact, which occurs when, soon after leaving the $s=0$ end, the loop passes close to its other end and the attached DNA segment (Figs.~\ref{fig:iter_short}, \ref{fig:aniso_drw}, \ref{fig:range_0}). The contact gets closer as the bending anisotropy increases or $A/C$ decreases (Fig.~\ref{fig:range_0}). If electrostatics were taken into account at this moment, this near self-contact would inflict a strong energetic penalty due to the strong self-repulsion of the negatively charged DNA. This happens indeed, as it will be demonstrated in the next section, and the open shape and the absence of any self-contact becomes an additional argument in favor of selecting the U solution as our prediction of the shape of the real DNA loop formed by the real \lr.


\subsection{Bending anisotropy effects for the 385~bp {\it lac} repressor loops}
\label{sec:anisotropy_range_long}

The described effects of bending anisotropy were similarly observed in the case of the 385~bp loops. Bending concentrated in the ``soft joints'' and the solutions developed oscillations of curvature, twist, and energy density. For $\mu=1/4$, the solutions became more bent and less twisted on the average, as the data in Table~\ref{tab:long_aniso} show (cf. Table~\ref{tab:long_sols}). The U solution was the one to undergo the least change from its isotropic shape, while the solutions O and \OP were those that changed the most.

\begin{table}[!t]
\begin{ruledtabular}
\begin{tabular}{ccccccc}
Solu- & $<\omega>$ & $<\kappa>$ & $\kappa_{max}$ & $U$ & \Rmsd \\
tion & (\dbp) & (\dbp) & (\dbp) & (\mbox{\scriptsize kT}) & (h) \\
\hline
U & -0.20 & 0.82 & 2.16 & 4.2 & 3.3 \\
O & 0.26  & 0.96 & 2.69 & 6.0 & 9.1 \\
\UP & -0.29 & 1.34 & 2.61 & 9.7 & 4.7 \\
\OP & 0.34 & 1.38 & 2.66 & 10.5 & 7.6 \\
\end{tabular}
\end{ruledtabular}
\caption{\label{tab:long_aniso}Geometric and energetic properties of the four solutions of the BVP problem for the 385~bp-long DNA loop, in the case of $\mu=1/4$ ($\alpha = 4/15$, $\beta = 16/15$). \Rmsd\,s are computed with respect to the isotropically flexible structures with $\alpha = \beta = 2/3$.}
\end{table}

In the broad range of parameters $\alpha$ and $\beta$, the four long loops showed the same tendencies as the two short loops.  The bending anisotropy allowed for a significant reduction in the elastic energies, making the loops effectively softer (more bendable).  The shapes of the loops, after undergoing some transformation following the change in the bending anisotropy or in the loop rigidity, eventually reached asymptotic states.  The asymptotic states for the soft loops (those with the small $A/C$ ratio and/or high $\mu$) were strongly bent conformations with practically zero unwinding/overwinding, where the twisting energy was of the same order as the small bending energy.  The asymptotic states for the rigid loops (those with the large $A/C$ ratio and $\mu$ on the order of 1) were the conformations with the least possible bending for each given loop topology, where the twisting achieved the worst possible value so as to optimized the bending, yet the latter still accounted for most of the elastic energy. Same as in the case of the short loops, the shapes of the overwound and underwound solutions converged when the loop rigidity achieved particularly large values.

\begin{figure}[tb]
\begin{center}
\includegraphics[scale=.43]{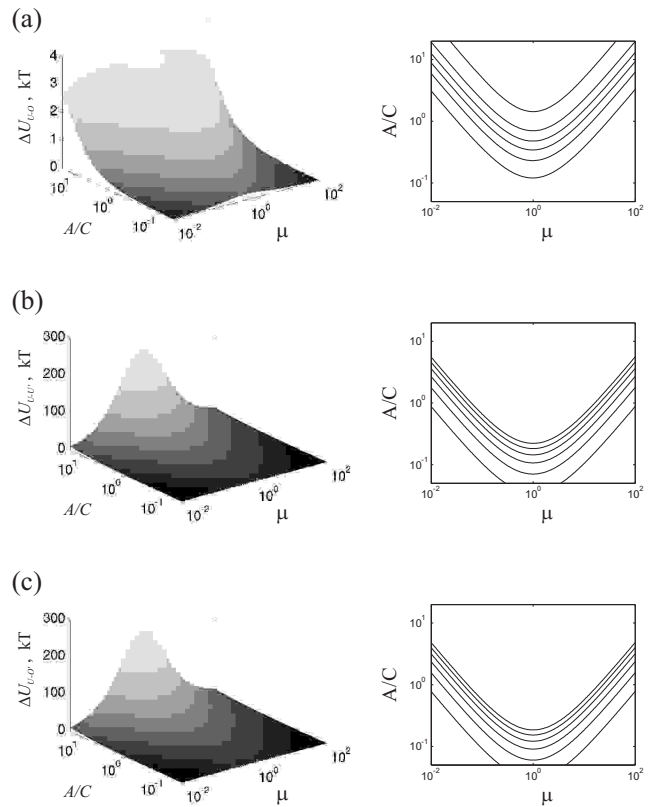}
\end{center}
\caption{\label{fig:uo_diff_long}Elastic energy difference between U and (a)~O, (b)~\UP, and (c)~\OP\ solutions for the 385~bp-long DNA loop.  3D plots of the energy difference in the coordinates $\mu = \alpha/\beta$ and $A/C \approx (\alpha+\beta)/2$~\eqref{eq:bendmod_tw} are shown on the left, and contour maps of the 3D plots for the $\Delta U$ values of 0.5, 1, 1.5, 2, 2.5, and 3\,kT are shown on the right.}
\end{figure}

The underwound U solution was again the one to have the least elastic energy among the four solutions throughout the whole studied range of $\alpha$ and $\beta$ ($A/C$ and $\mu$) values. The maps of the energy difference between U and the other three solutions are shown in Fig.~\ref{fig:uo_diff_long}.  The energy of the O solution does not normally differ from that of the U solution by more than several kT, therefore the O solution should contribute to a small extent to the thermodynamic ensemble of the loop shapes and the properties of the \lr-bound 385~bp DNA loop.  In contrast, the energies of the \UP\ and \OP\ loops are consistently 2-2.5 times larger than the energy of the U loop -- and the difference amounts to small kT values only for unlikely combinations of $A/C$ and $\mu$.  Therefore, one can safely conclude that these two loops, even though uninhibited by steric overlap with the \lr, are still extraneous solutions, as in the case of the 76~bp loop, and may be safely excluded from any computation involving the properties of the thermodynamic ensemble of the 385~bp loop conformations.


\section{Electrostatic effects}
\label{sec:electrostatics}

The last, but perhaps, the most important extension of the classic theory that is included in our model, consists of ``charging'' the modeled DNA molecule.  The phosphate groups of a real DNA carry a substantial electric charge: $-20.8e$ per helical turn, that significantly influences the conformational properties of DNA~\cite{SCHL94,BEDN94,MANN78}.  The DNA experiences strong self-repulsion that stiffens the helix and increases the distance of separation at the points of near self-contact.  Also, all electrostatically charged objects in the vicinity of a DNA -- such as amino acids of an attached protein, or lipids of a nearby nuclear membrane -- interact with the DNA charges and influence the DNA conformations.  Below, we describe our model of the electrostatic properties of DNA and the effects of electrostatics on the conformation of the \lr\ DNA loops.


\subsection{Changes to the equations of elasticity due to electrostatics}

The electrostatic interactions of DNA with itself and any surrounding charges are introduced in our theory through the body forces $\vec{f}$:
\begin{equation}
\dot{\vec{f}}(s) = \sigma(s) \E(\vec{r}(s))\:,
\label{eq:f_elec}
\end{equation}
where $\E$ is the electric field at the point $\rv(s)$ and $\sigma(s)$ is the density of DNA electric charge at the point $s$.  The present simplified treatment places the DNA charges on the centerline, as it was done in other studies~\cite{WEST97}.  Implications of a more realistic model will be discussed in Sec.~\ref{sec:discussion}.

The charge density $\sigma(s)$ is modeled by a smooth differentiable function with relatively sharp maxima between the DNA base pairs, where the phosphate charges should be located.  The chosen (dimensionless) function:
\begin{equation}
\sigma(s) = \frac{8}{3} \QD \sin^4 \left( \pi \ND s \right)
\label{eq:sigma}
\end{equation}
\noindent is somewhat arbitrary but specifics are unlikely to significantly influence the results of our computations, as will be discussed below. $\ND$ denotes the number of base pairs in the modeled DNA loop (which is assumed to be starting and ending with a base pair) and $\QD$ denotes the total charge of that DNA loop. That charge is reduced from its regular value of $2e$ per base pair due to Manning counterion condensation around the phosphates~\cite{MANN78}: $\QD = 2e\chi\ND$.  In this work, we assume $\chi = 0.25$, which is an observed value for a broad range of salt concentrations~\cite{MANN78}.

\begin{figure}[tb]
\begin{center}
\includegraphics[scale=.35]{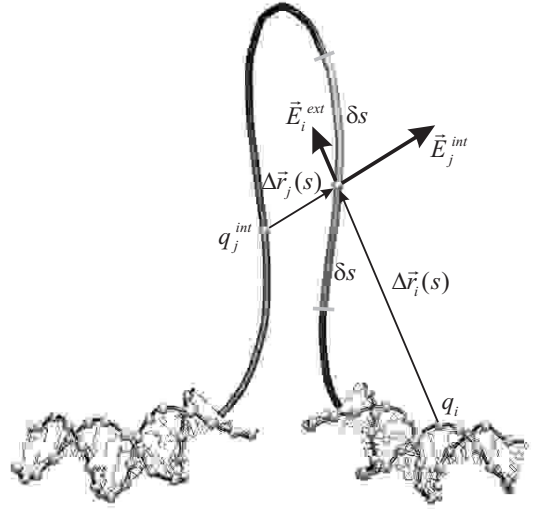}
\end{center}
\caption{\label{fig:el_setup}Electrostatic interactions in the elastic rod problem. The electric field $\vec{E}(s)$ is computed at each point $s$ of the rod as the sum of the ``external'' field $\vec{E}^{ext}$, produced by the charges $q_i$, not associated with the elastic rod, and the ``internal'' field $\vec{E}^{int}$, produced by the charges $q^{int}_j = 2e\chi$, placed in the maxima of the charge density $\sigma(s)$ of the elastic rod~\eqref{eq:sigma}. Here, $\Delta \vec{r}_i(s) = \vec{r}(s) - \vec{R}_i $ and $\Delta \vec{r}_j(s) = \vec{r}(s)-\vec{r}(s_j)$; see eq.~\eqref{eq:E_field} for detail. The area of the rod shown in light color lies within the cutoff $\delta s$ and is excluded from computations of the electric field at the point $s$.}
\end{figure}

The electric field $\E$ is composed of the field of external electric charges, not associated with the modeled elastic rod -- whichever are included in the model~-- and from the field of the modeled DNA itself (Fig.~\ref{fig:el_setup}). $\E$ is computed using the Debye screening formula:
\begin{multline}
\E(\vec{r}(s)) = {1 \over 4\pi\epsilon\epsilon_\circ} \; \Bigg( \sum_i q_i \vec{\nabla}\; {\exp(-|\vec{r}(s)-\vec{R}_i|/\lambda) \over |\vec{r}(s)-\vec{R}_i|} \\
+ 2 e \chi \sideset{}{'}{\sum}_j \vec{\nabla}\; {\exp(-|\vec{r}(s)-\vec{r}(s_j)|/\lambda) \over |\vec{r}(s)-\vec{r}(s_j)|} \Bigg) \:,
\label{eq:E_field}
\end{multline}

\noindent where $\lambda = 3\,{\rm \AA}/\sqrt{c_s}$ is the radius of Debye screening in an aqueous solution of mono valent electrolyte of molar concentration $c_s$ at 25 $^\circ$C~\cite{MANN78}. The dielectric permittivity $\epsilon$ is that of water: $\epsilon = 80$.  

The first term in eq.~\eqref{eq:E_field} represents the DNA interaction with external charges $q_i$, located at the points $\vec{R}_i$; the sum runs over all those charges.  The second term represents the self-repulsion of the DNA loop, and that sum runs over all the maxima of the charge density $\sigma(s)$, where the DNA phosphates -- charges of $2e\chi$ -- are located. This sum approximates the integral over the charged elastic rod; computing such integral would be more consistent with the chosen model. However, this approximation is rather accurate, as will be shown below, and is made in order to significantly reduce the amount of computations required to calculate the electric field.

More importantly, certain phosphate charges are excluded from the summation in the second term (hence the prime sign next to the sum). Those excluded are the charges that are located closer to the point $s$ than a certain cutoff distance $\delta s$ (Fig.~\ref{fig:el_setup}).  The reason for introducing such cutoff is that the DNA elasticity has partially electrostatic origin, so that the energetic penalty for DNA bending and twisting, approximated by the elastic functional~\eqref{eq:ener_funct}, already includes the contribution from electrostatic repulsion between the neighboring DNA charges. It is debatable what ``neighboring'' implies here, i.e., how close should two DNA phosphates be in order to significantly contribute to DNA elasticity. In this work, the cutoff distance $\delta s$ is chosen to be equal to one step of the DNA helix ($h$=36\,\AA). This, on the one hand, is the size of the smallest structural unit of DNA, beyond which it does not make sense at all to use a continuum model of the double helix -- so, at the very least, the phosphate pairs within such unit should be excluded from the explicit electrostatic component.  On the other hand, the forces of interaction between the phosphates, separated by more than that distance from each other, are already much smaller than the elastic force, as will be shown below.  Thus, even though the chosen cutoff $\delta s$ might be too small, the resulting concomitant stiffening of the DNA is negligible. For comparison, calculations with cutoff values $\delta s = 1.5\,h$ and $\delta s = 2\,h$ were also performed. 

Thus, the electric field $\vec{E}$, computed using~\eqref{eq:E_field}, is substituted into~\eqref{eq:f_elec}, and the resulting body forces $\dot{\vec{f}}$ appear in Eqs.~(\ref{eq:adv_grand_1}), (\ref{eq:adv_grand_2}), (\ref{eq:adv_grand_4}) of the ``grand'' system, in place of the previously zeroed terms.  ``Unzeroing'' those terms, however, is not the only change to the equations.  More importantly, these body force terms depend on the conformation of the entire elastic loop due to the self-repulsion term in~\eqref{eq:E_field}.  This makes the previously ordinary differential equations of elasticity integro-differential and therefore, requires a new algorithm for solving them. The solutions of the integro-differential equations minimize the new energy functional:
\begin{equation}
U = U_{elastic} + U_{Q} - U_{Q,straight}\:,
\label{eq:ener_funct_elec}
\end{equation}
\noindent where $U_{elastic}$ is the elastic energy computed as in~\eqref{eq:ener_funct}, $U_{Q}$ is the electrostatic energy computed, in accordance with~\eqref{eq:E_field}, as
\begin{multline}
U_{Q} =  {1 \over 4\pi\epsilon\epsilon_\circ} \; \int_0^1 \; \sigma(s) \Bigg( \sum_i q_i {\exp(-|\vec{r}(s)-\vec{R}_i|/\lambda) \over |\vec{r}(s)-\vec{R}_i|} \\
+ 2 e \chi \sideset{}{'}{\sum}_j {\exp(-|\vec{r}(s)-\vec{r}(s_j)|/\lambda) \over |\vec{r}(s)-\vec{r}(s_j)|} \Bigg) ds \:,
\label{eq:elec_ener}
\end{multline}
and $U_{Q,straight}$ is the electrostatic ``ground state'' energy: computed using the same formula~\eqref{eq:elec_ener} for the straight DNA segment of the same length as the studied loop.


\subsection{Changes to the computational algorithm; results for the 76~bp lac repressor loops}

To solve the integro-differential equations, the following algorithm is employed.  As previously, the electrostatic interactions are ``turned on'' during a separate iteration cycle. At each step of the latter, the equations are solved for the electric field $\E_i = \wE \E$, where the ``electrostatic weight'' $\wE$\ grows linearly from 0 to 1.

\begin{figure}[!t]
\begin{center}
\includegraphics[scale=.45]{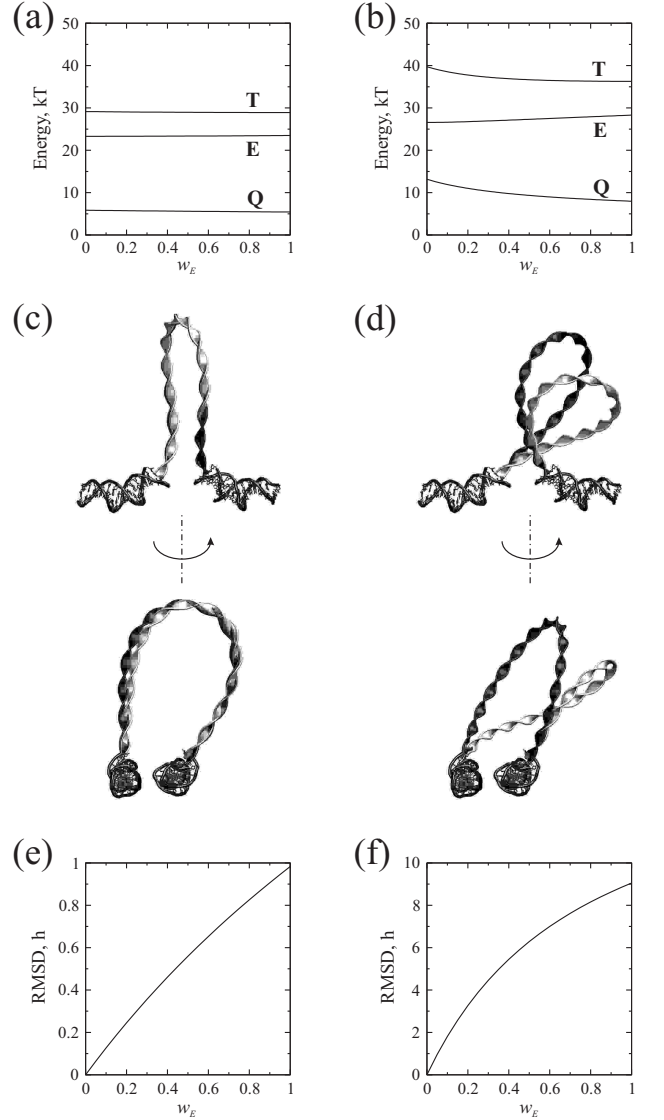}
\end{center}
\caption{\label{fig:el_eff}Changes in the predicted structure and energy of the 76~bp DNA loops after electrostatic interactions are taken into account. Left column: the U solution; right column: the O solution. (a,b) Elastic (E), electrostatic (Q), and total (T) energy of the elastic loop vs. the weight $w$ of the electrostatic interactions.  (c,d) Uncharged ($w=0$, in light color) and completely charged ($w=1$, in dark color) structures of the elastic loop. The bottom views are rotated by 70$^\circ$ around the vertical axis relative to the top views. (e,f) The \rmsd\ between the charged and uncharged loop structures, in DNA helical steps h.  The data in this figure correspond to $\alpha = 4/15$, $\beta = 16/15$, ionic strength of 10\,mM and an electrostatics exclusion radius $\delta s = h$.}
\end{figure}

However, each step of this iteration cycle becomes its own iterative sub-cycle. The electric field $\E_i$ is computed at the beginning of the sub-cycle and the equations are solved with this, constant field. Then the field is re-computed for the new conformation of the elastic rod, the equations are solved again for the new field, and so on until convergence of the rod to a permanent conformation (and, consequently, of the field to a permanent value). The weight $\wE$ is kept constant throughout the sub-cycle. To enforce convergence, the field used in each round of the sub-cycle is weight-averaged with that used in the previous round:
\begin{equation}
\E_{i,j} = w_a\; \E_{i,j(actual)} + (1-w_a) \E_{i,j-1}
\end{equation}
\noindent The averaging weight $w_a$ is selected by trial and error so as to speed up convergence. For the \lr\ system, the trivial choice of $w_a=0.5$ turned out to be satisfactory.

This approach to solving the integro-differential equations assumes that the elastic rod conformation changes smoothly with the growth of the electric field.  For intricate rod conformations, which might change in a complicated manner with the addition of even small electrostatic forces, this approach may conceivably fail.  Yet, it worked extremely well for the studied case of the DNA loop clamped by the \lr.

The changes to the structure and energy of the 76~bp DNA loops due to electrostatic interactions were computed for a broad range of ionic strength $c_s$ (0, 10, 15, 20, 25, 50, and 100\,mM) and three different cutoff values $\delta s$ ($h$, $1.5\,h$, $2\,h$). The computations were performed with $A/C = 2/3$ and the previously selected $\mu = 1/4$ (resulting in the elastic moduli of $\alpha = 4/15$ and $\beta = 16/15$). The external charges included in the model were those associated with the phosphates of the DNA segments from the crystal structure~\cite{LEWI96} (see Fig.~\ref{fig:el_setup}). The iteration cycle was divided into 100 sub-cycles, which showed a remarkable convergence: the length of no sub-cycle exceeded three iteration rounds.

The changes in the structure and energy of the elastic loops due to electrostatic interactions are presented in Fig.~\ref{fig:el_eff} for the ionic strength of 10\,mM and the exclusion radius of $h$. The structure of the U solution almost does not change: the \rmsd\ between the original ($\wE = 0$) and the final ($\wE = 1$) structures does not exceed $1\,h$.  Neither do the curvature and twist profiles of this loop significantly change (Fig.~\ref{fig:short_plots}, 3d column). The energy of the loop changes by the electrostatic contribution of 6.1\,kT, that -- because the structure is not changing -- practically does not depend on $\wE$ (Fig.~\ref{fig:el_eff}). This energy increase is mainly accounted for by the interaction of the loop termini with each other and the external DNA segments (Fig.~\ref{fig:short_plots}). The self-repulsion accounts for 55\%\ of the electrostatic energy; 45\%\ comes from the interaction with the external DNA segments. The apparent reason for the absence of a large change in the geometry of the U loop lies in the fact that this loop is an open semicircular structure, which at no point comes into close contact with itself or the external DNA. 

In contrast, the initial structure of the O loop is bent over so that the beginning of the loop almost touches the end of the loop and the attached DNA segment (Fig.~\ref{fig:aniso_drw}, \ref{fig:el_eff}\,d).  As a result, the electrostatic interactions force a significant change in the structure and energy (Fig.~\ref{fig:el_eff}\,b,\,d,\,f).  The structure opens up, the gap at the point of the near self-crossing increases, the \rmsd\ between the final and the original structures reaches $9\,h$ (Fig.~\ref{fig:el_eff}\,f); the DNA overwinding almost doubles (Fig.~\ref{fig:short_plots}, 6th column). This allows the electrostatic energy to drop from 13.2\,kT to 8.0\,kT, yet the elastic energy grows by 1.7\,kT (Fig.~\ref{fig:el_eff}\,b); together, the energy reaches 36.3\,kT so that the gap from the U loop increases from 3.3\,kT to 7.4\,kT. As in the case of the U loop, the main contribution to the electrostatic interactions comes from the loop ends (Fig.~\ref{fig:short_plots}); the energy distribution between self-repulsion and the interactions with the external DNA charges is practically the same.

\begin{figure}[!t]
\begin{center}
\includegraphics[scale=.145]{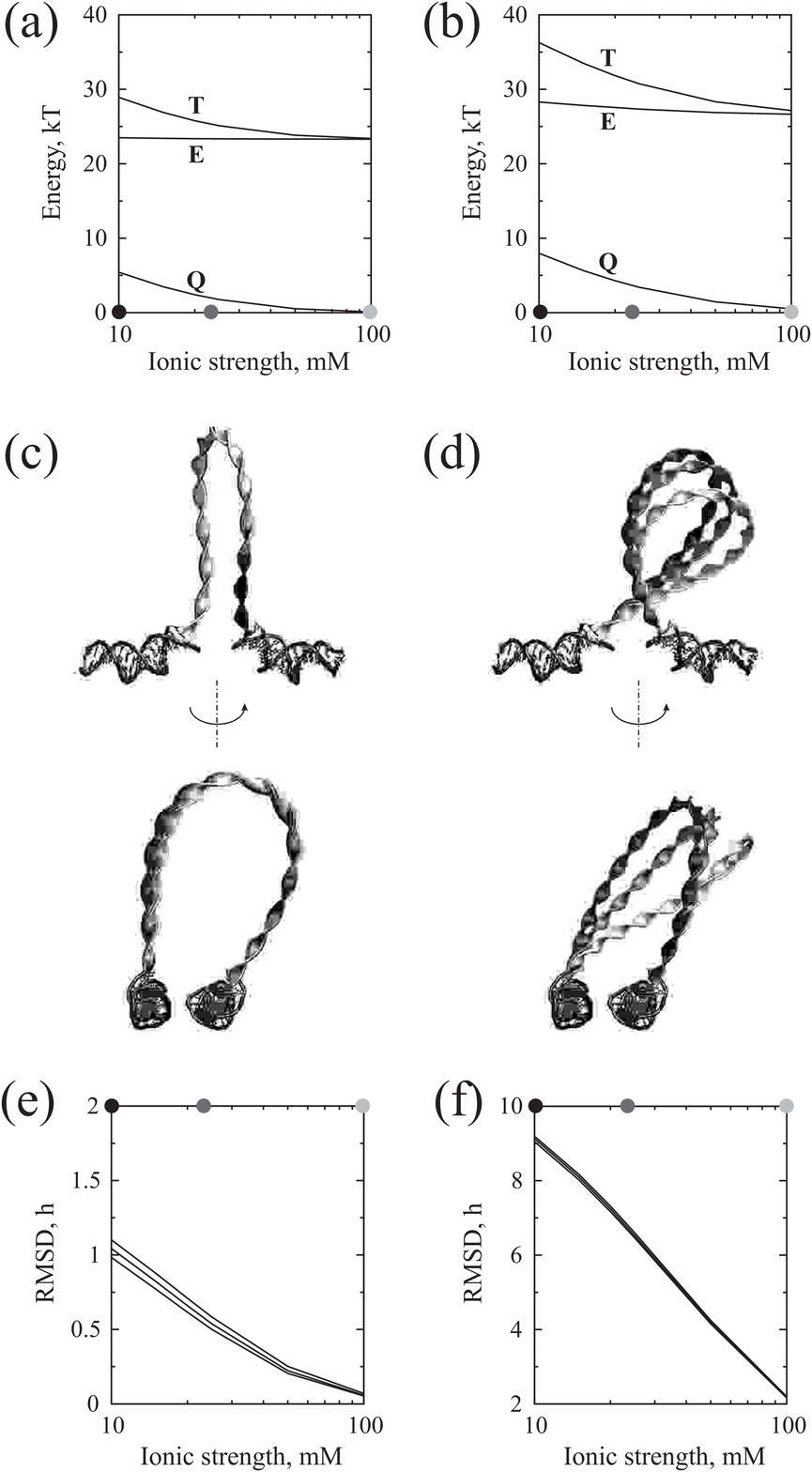}
\end{center}
\caption{\label{fig:ion_change}The effect of ionic strength on the predicted structure and energy of the 76~bp DNA loops. Left column: the U solution; right column: the O solution. (a,b) Elastic (E), electrostatic (Q), and total (T) energy of the elastic loop.  Shown are only the plots for the exclusion radius of $h$; the energy plots for the other exclusion radii are almost indistinguishable. (c,d) Snapshots of the elastic loop structures for 10\,mM (dark), 25\,mM (medium), and 100\,mM (light color). Points where the snapshots were taken are shown as dots of the corresponding colors on the axes of panels (a),~(b),~(e),~(f). The bottom views are rotated by 70$^\circ$ around the vertical axis relative to the top views. (e,f) The \rmsd\ of the loop structures from those computed without electrostatics (equivalent to infinitely high salt). The lines, from top to bottom, correspond to the exclusion radii of $h$, $1.5\,h$, and $2\,h$.}
\end{figure}

\begin{figure*}[!t]
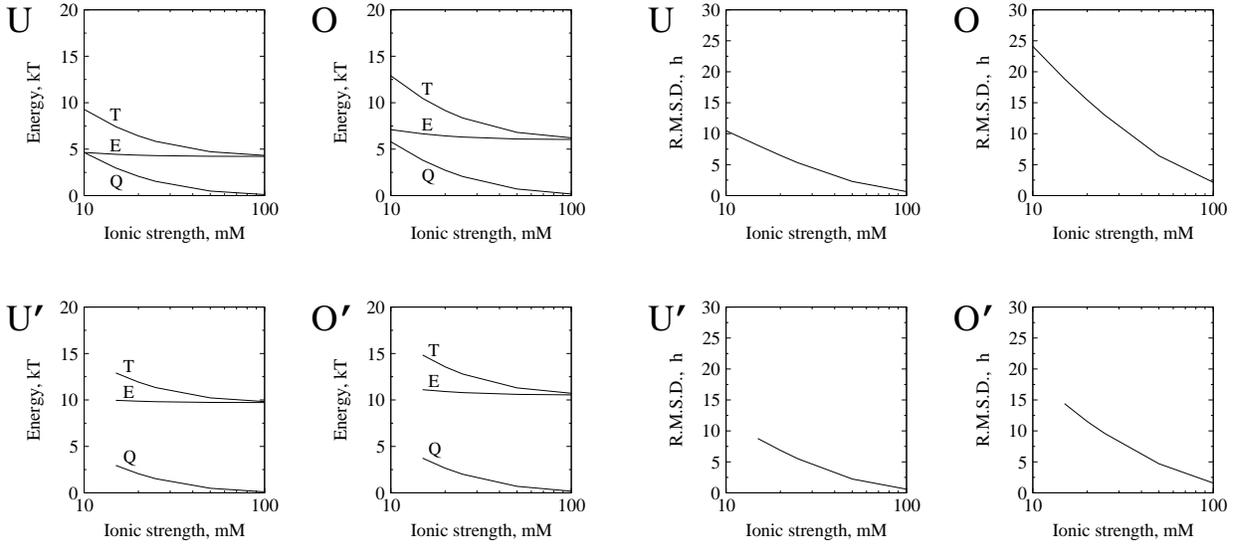

\begin{center}
\includegraphics[scale=.4]{Fig_Ion_str_Long_E_plot}\hspace{8mm}\includegraphics[scale=.4]{Fig_Ion_str_Long_RMSD_plot}
\end{center}
\caption{\label{fig:ion_change_long}The effect of ionic strength on the energy and geometry of the 385~bp DNA loops. Elastic (E), electrostatic (Q), and total (T) energy (left panel) and \rmsd\ from the uncharged structures (right panel) are plotted for the U, \UP, O, and \OP\ loops vs the ionic strength $c_s$ for the exclusion radius of $3\,h$. The missing points on the \UP\ and \OP\ plots indicate solutions, missing due to the non-convergence of the iterative procedure (see text).}
\end{figure*}

Naturally, the calculated effect diminishes when the ionic strength of the solution increases and therefore the electrostatics becomes more strongly screened.  Fig.~\ref{fig:ion_change} shows what happens to the structure and energy of the U and O loops when the ionic strength changes in the range of 10\,mM -- 100\,mM (which covers the range of physiological ionic strengths). The structure and elastic energy of the U loop again show almost no change, the total energy of that loop falls from 29.7\,kT to 23.5\,kT due to the drop in electrostatic energy.  The structure of the O loop returns to almost what it was before the electrostatics was computed (within the \rmsd\ of $2.2\,h$); the elastic energy of this loop drops back to 26.6\,kT and the electrostatic energy -- to mere 0.5\,kT. This results show that theoretical estimates of the energy of a DNA loop formation in vivo need to employ as good an estimate of the ionic strength conditions as possible.

The \lr\ loops were extensively used to analyze all the assumptions and approximations of our model and showed that those were satisfactory indeed. The calculations were repeated for the self-repulsion cutoffs of $\delta s = 1.5\,h$ and $\delta s = 2\,h$.  The resulting change in the loop energy at $c_s=$10\,mM equals only $\Delta\,U_{h-1.5\,h} = 0.35\,{\rm kT}$ and $\Delta\,U_{h-2\,h} = 0.75\,{\rm kT}$ for the U loop, and $\Delta\,U_{h-1.5\,h} = 0.35\,{\rm kT}$ and $\Delta\,U_{h-2\,h} = 0.75\,{\rm kT}$ for the O loop; all these values drop to below 0.1\,kT when the ionic strength rises to 100\,mM. The difference is mainly in electrostatic energy, and the elastic energy is always within 0.1\,kT of that of the structures obtained with $\delta s = h$.  Accordingly, the \rmsd\ from the uncharged structure varies by at most $0.1\,h$ for the different cutoffs (Fig.~\ref{fig:ion_change}\,e,\,f).  Therefore, even the ``largest'' cutoff $\delta s = 2\,h$ is satisfactory for the electrostatic calculations -- while at the same time increasing the speed of computations.

An additional advantage of using larger cutoff is that the concomitant stiffening of the modeled DNA, which possibly takes place due to too many phosphate pairs included in the electrostatic interactions, is reduced.  Such stiffening, however, is truly negligible: in all the cases, the electrostatic force does not exceed 1-2\,pN per base pair, compared to the calculated elastic force in the range of 10-20\,pN.  Changes in the cutoff results in only insignificant changes of the electrostatic force. Nor does evaluating the electric field and energy using the sums~\eqref{eq:E_field} and \eqref{eq:elec_ener} (instead of a more consistent integral over the loop centerline) result in any significant difference.  Test calculations showed that in all the studied cases the two ways of evaluating the energy differ by at most 0.02\%.

Finally, it was tested in how far the particular choice~\eqref{eq:sigma} for the charge density of DNA $\sigma(s)$ influences the computation results.  The calculations were repeated for the constant charge density $\sigma^\prime(s) = \QD$ (in dimensionless representation). The energies of the loop conformation never changed by more than $10^{-4}$ of their values in the whole range of $c_s$ and $\delta s$; the \rmsd\ between the loop conformations obtained with different $\sigma(s)$ never exceeded $0.01\,h$.  Therefore, the electrostatic properties of the elastic rod in the current model can safely be computed with constant electrostatic density, further saving the computation time.


\subsection{Results for the 385~bp lac repressor loops}
\label{sec:electro_385}

The electrostatic computations were similarly performed for the 385~bp loops, in the same range of salt concentrations and for exclusion radii of $2\,h$ and $3\,h$. For the U and O loops, the results were qualitatively the same as in the case of the short loops. The elastic loops became more open and straigtened up with respect to the \lr; the energy of the loops increased by 0-6\,kT, depending on the salt concentration (Fig.~\ref{fig:ion_change_long}). The U loop was again the one to change its structure and energy to the least extent. The results of the computations using the different cutoff radii were practically the same; approximating the self-repulsion field by a discrete sum~\eqref{eq:E_field} gave almost exact results; replacing the charge density function~\eqref{eq:sigma} with the constant function $\sigma^\prime(s)$ had no significant influence on the results.

One difference from the short loop case consisted, though, in the more significant change of the long loop structures with respect to the uncharged loops. The \rmsd\,s reached $10\,h$ for the U loop and $25\,h$ for the O loop (Fig.~\ref{fig:ion_change_long}, cf Fig.~\ref{fig:ion_change}\,e,\,f). As previously, the major part of the electrostatic effect consisted in repulsion between the ends of the loop, brought closely together by the protein, and the protein-attached DNA segments. This repulsion tended to change the direction of the ends of the loop, bending them away from each other and the DNA segments.  In the case of the short loops it was impossible to notably change the direction without significantly stressing the rest of the loop. Yet the long loops could more easily accomodate opening up at the ends and therefore changed their structures more significantly. 

The larger structural change necessitated longer calculations. For the long loops, the iteration steps typically consisted of 5-6 iteration sub-cycles, and even of a few dozen sub-cycles at especially stiff steps.

The \UP\ and \OP\ loops showed a similar responce to the electrostatics at high salt concentration (above 25\,mM).  Their electrostatic energy lied in the range 0-5\,kT, and the structural change due to the increased bending of the loop ends amounted to up to $10\,h$\ \rmsd\ from the uncharged structure for the \UP\ solution and up to $15\,h$ -- for the \OP\ solution (Fig.~\ref{fig:ion_change_long}). Yet, low salt and stronger electrostatics rendered the solutions instable. Electrostatic computations with no salt screening (0\,mM) transformed the \UP\ solution into the U solution and the \OP\ solution -- into the O solution. What made the solutions instable was apparently the ever increasing bending of the ends of the loops away from each other that, in combination with the bending anisotropy, also caused high twist oscillations near the loop ends (as described in Sec.~\ref{sec:anisotropy_structure}). The combination of twisting and bending caused the loops to flip up -- as one can cause a piece of wire to flip up and down by twisting its ends between one's fingers.

For the intermediate salt concentrations (10-20\,mM) the oscillations of the intermediate solutions between the two possible states resulted in non-convergence of the iterative procedure. In this respect, using the larger electrostatic exclusion radius improved convergence. For the exclusion radius of $2\,h$, the iterations did not converge for salt concentrations of 15 and 20\,mM; convergence for the 10\,mM salt was achieved but resulted again in flipping up to the stable solutions. For the exclusion radius of $3\,h$, the iterations successfully converged to \UP\ and \OP\ solutions (albeit somewhat changed due to the electrostatics) for the 15 and 20\,mM salt concentrations and did not converge for 10\,mM only.

Such instability of the \UP\ and \OP\ loops serves, of course, as another argument for disregarding them in favor of the stable U and O solutions.

Upon introducing the electrostatic self-repulsion, an interesting experiment could be performed. Self-crossing by the solutions during the iteration cycles -- as described in Sec.~\ref{sec:lacsols_short} -- was no longer possible. Therefore, one could explore whether superhelical structures of the loop could be built by further twisting the ends of the loop. One extra turn around the cross-section at the $s=1$ end did generate new structures of the U and O loops. Yet, those structures were so stressed and had such a high energy (on the order of 50\,kT higher than their predecessors) that it was obvious that those structures will not play any part at all in the real thermodynamic ensemble ot the \lr\ loops. Any further twisting of the ends resulted in non-convergence of the iterative procedure. Clearly, this length of a DNA loop is insufficient to produce a rich spectrum of superhelical structures.


\section{Discussion}
\label{sec:discussion}

Below, we will review the presented modeling method and its limitations, compare it with the previously reported similar models, discuss the possible applications and the further developments of the method, and summarize what has been learned about the \lr-DNA complex.


\subsection{Summary of the model}

The presented modeling method consists in approximating DNA loops with electrostatically charged elastic rods and computing their equilibrium conformations by solving the modified Kirchhoff equations of elasticity. The solutions to these equations provide zero-temperature structures of DNA loops  -- the equilibrium points, around which the loops fluctuate at a finite temperature. From these solutions, one can automatically obtain both global and local structural parameters, such as the twist and curvature at every point of the loop, the loop's radius of gyration, the linking number of the loop and its distribution between writhe and twist, various protein-DNA distances, the distances between DNA sites of special interest, etc. The solutions readily provide an estimate of the energy of the DNA loop and the forces that the protein has to muster in order to confine the loop termini to the given conformation, the distribution of the energy between bending and twisting, and the profiles of stress and energy along the DNA loop.

Our method allows to build a family of topologically different loop structures by applying such simple geometrical transformations as twisting and rotating the loop ends, or varying the initial loop conformation that serves as the starting point for solving the BVP. By comparing the energies of the topologically different conformations and assuming the Boltzmann probability to find the real DNA loop in either of them, one can either deduce the lowest energy structure which the loop should predominantly adopt -- as was mostly the case in the studied \lr-DNA complex -- or to compute the loop properties of interest by taking Boltzmann averages among several obtained structures of comparable energies.

For DNA loops of a size of several persistence lengths (150~bp~\cite{HAGE88,BAUM97}), only a few topologically different structures of comparable energy can exist and our simplified search of the conformational space should be sufficient to discover all members of the topological ensemble -- as it has been demonstrated in this work. Yet longer DNA loops should produce large topological families of structures and, unless a fast exhaustive search procedure is discovered, this limits the applicability of our method to DNA loop on the order of or shorter than 1,000~bp.  The method is still good for generating sample structures of longer loops, but more structures of comparable energy are likely to be missed.

The lower boundary of applicability of our method is stipulated by the loop diameter: the Kirchhoff theory is applicable only if the elastic rod is much longer than its diameter.  The diameter of the DNA double helix is 2~nm, which is equivalent to 6~bp. Therefore, the loop studied by our method should be at least several times longer than that: roughly, 50~bp and longer.  Ideally, one would request that the loop exceed the persistence length of DNA, yet proteins are known to bend DNA on smaller scales -- as, for example, the \lr\ does -- so we apply the theory to DNA loops of around 100~bp in length.

Hence, we suggest that the presented model can be applied to studying the conformations of DNA loops of about 100~bp to 1,000~bp length.  Building a single conformation is a fast process that takes only several hours of computation on a single workstation. The problem can be solved for a certain set of boundary conditions, as presented here -- or the loop boundaries can be systematically moved and rotated and the dependence of the loop properties on the boundary conditions can be studied by re-solving the problem in each new case. Such approach has the advantage over the Monte Carlo simulations in avoiding having to build and analyze a massive set of structures sampling the conformational space. 

At a finite temperature, the DNA loops exist as an ensemble of conformations. While generating all feasible topologically different structures, our method still neglects the thermal vibrations of each of them and the related entropic effects. Yet, those effects are likely to be insignificant, since the length of the studied DNA loops is limited to several persistence lengths, as discussed above. The related thermal oscillations of the loop structure should be small, although a separate study to quantify the effect of the oscillations seems worthwhile. For longer DNA loops, an extensive sampling of the conformational space -- for example, using the Monte Carlo approach -- is indispensable.


\subsection{Advanced features and potential applications}
\label{sec:discussionB}

Compared to the pre-existing analytical and computational DNA models based on the theory of elasticity~\cite{MARK94,MARK95A,MARK98,SHI95,COLE95,WEST97,KATR97,TOBI2000,COLE2000,MATS2002}, our model provides a universal and flexible description of DNA properties and interactions.  Most previous methods either considered the DNA to be isotropically flexible~\cite{MARK94,MARK95A,MARK98,SHI95,WEST97,KATR97,TOBI2000,COLE2000}, or did not consider the effects of the DNA intrinsic twist and curvature~\cite{COLE95}, or limited the treatment to a purely elastic model, that is, to the cases when the electrostatic properties of DNA could be disregarded~\cite{MARK94,MARK98,MATS2002}.  In the present work, a model of anisotropically flexible, electrostatically charged DNA with intrinsic twist and intrinsic curvature has been employed,  Kirchhoff equations were derived in their most general form~\refgrandA, and all these DNA properties have been extensively studied in the case of the \lr\ loops.  As it has been shown in Sec.~\ref{sec:anisotropy}, the combination of the intrinsic twist with the anisotropic flexibility is essential in order to correctly estimate the energy of the DNA loop, as well as the local bend and twist at each point of the loop.  The electrostatic interactions are important in the case of a close contact of the loop with itself or with other molecules (Sec.~\ref{sec:electrostatics}).  The universality of our approach allows us to include all these cases into the scope of approachable problems.

Moreover, all the DNA parameters: the bending moduli, the intrinsic twist and curvature, the charge density -- are treated in our equations as functions of the arclength. This provides an automatic tool for studying the sequence-dependent properties of the DNA loops.  The elastic moduli and intrinsic parameters need to be specified for different DNA sequences, e.g., as outlined in~\cite{MATS2002,OLSO98,HOGA87}.  Then the properties of any loop are approximated by functions $\alpha(s)$, $\beta(s)$, $\gamma(s)$, $\kiot(s)$, $\omo(s)$, smoothly changing their values between those associated with each base pair in the loop sequence (as discussed in App.~\ref{sec:appendixd}) and the influence of the DNA sequence on the conformation and energy of that loop can be studied by simply substituting their functions associated with different DNA sequences into Eqs.~\refgrandA. Of course, extensive testing of such parameters and comparison with experimental results on DNA bending and loop energetics has to be done prior to performing any such studies of sequence-dependent effects.

Another opportunity that our model provides consists in its ability to mimic the effect of protein binding within DNA loops. For example, if a protein is known to bend DNA over a certain region~\footnote{So does a huge variety of DNA-binding proteins~\cite{LILL95}, for example, CAP~\cite{BUSB99}, TATA-binding protein~\cite{LILL95}, or, indeed, the headgroup of the \lr\ itself~\cite{LEWI96}.} then the intrinsic curvature term can be adjusted so as to enforce the required curvature over that region of the DNA loop. To strictly enforce the curvature, the rigidity of that region would have to be increased as well.  Studying the solutions to the Kirchhoff equations with such intrinsic curvature term would reveal what changes to the energy and conformation occur in the DNA loop upon binding of the specified protein. A recent study~\cite{BALA2000,BALA2003A} of the CAP protein binding within the \lr\ loop, performed in this vein, suggested an explanation for the experimentally observed cooperation between the CAP and the \lr\ in DNA binding~\cite{HUDS90,PERR96}.

Certainly, the different features of our model have to be used only as each specific problems dictates.  For example, the model of isotropically flexible DNA seems sufficient for determining the global properties of DNA loops, such as the linking numbers of different conformations, their radii of gyration, or the energy distribution between bend and twist.  In the present study, increasing the complexity of the model and varying the elasticity properties did not much change the shape of the loop, which determines these global properties (cf. Figs.~\ref{fig:iter_short}, \ref{fig:long_sols}-\ref{fig:range_0}, \ref{fig:el_eff}, \ref{fig:ion_change}). Presumably, the reduced model should be sufficient for computing the global properties of other DNA loops.

If, however, the energy of the loop has to be estimated, or the energies of different loop conformations need to be compared, or, indeed, the local structure at a certain point of the loop needs to be predicted, with a view to study how binding of a certain protein in that area changes upon the loop formation -- then the anisotropic loop model becomes imperative, as it has been demonstrated in Sec.~\ref{sec:anisotropy}. Finally, the problems which involve loop conformations with a near self-contact -- such as those of the O loop or the tightly wound superhelical structures~\cite{WEST97,YANG95,SCHL94} -- necessitate including the electrostatic force term into Kirchhoff equations. The same holds for the problems studying the effect of salt concentration on structure and energetics of DNA loops.

The speed and high adaptability of our modeling approach makes it a good candidate for multi-scale modeling simulations. The most obvious application would be to study the structure and dynamics of a system similar to the \lr-DNA complex: a protein or a protein aggregate holding a long DNA loop. The structure of the loop and the forces and torques that the loop exerts on the protein are directly obtainable from the elastic rod computations. Such forces would normally consist of the elastic forces at the loop boundaries and, perhaps, the electrostatic forces if parts of the DNA loop approach the protein-DNA complex closely. The electrostatic forces are directly computable from the predicted loop geometry. The forces and torques can then be plugged in an all-atom simulation of the structure of the protein aggregate and the DNA segments directly bound to it, similarly to how it is done in steered molecular dynamics simulations~\cite{ISRA2001,LECK2001}. The state of the DNA segments that provide the boundaries of the DNA loop can be routinely checked during the all-atom simulations. Whenever a notable change in the positions and orientations of the boundaries occur, the coarse-grained elastic computations have to be repeated (using the previous solutions as the initial guess) and the simulations continue with the updated values of the forces and torques. The fast coarse-grained computations presented here result in only a marginal increase in the total time required for the all-atom simulation. The multi-scale simulations provide a direct means to study the changes in structure and dynamics of the the protein aggregate under the force of the DNA loop it creates.

Another application of multi-scale modeling could be to a system where several proteins bind to the same DNA loop and alter the DNA geometry at or near their binding sites. Such systems arise during gene transcription, when multiple transcription factors and RNA polymerase components bind at or near a gene promoter, during DNA replication when helicases and gyrases concurrently alter the DNA topology, inside eukaryotic chromatin with multiple nucleosomes clustering on long DNA threads~\cite{BERG2002}. Simulations of the separate protein-DNA complexes could be run on atomic scale, and the stress imposed on the DNA in one simulation could be passed to another via an elastic rod model of the DNA segment(s) connecting the different complexes.  The binding of some proteins could even be mimicked with the intrinsic curvature and twist terms, as outlined above.

Finally, the local DNA geometry predicted in our model can be used to replace the coarse-grained elastic ribbon with an all-atom DNA structure in selected areas of the loop or even over the whole loop (cf App.~\ref{sec:appendixf}). An example is shown in  Fig.~\ref{fig:lac_operon}\,c,\,d, where the all-atom structure was placed on top of the predicted 76~bp and 385~bp U loops. A local segment of the all-atom structure can be simulated with the forces and torques from the rest of the loop applied to its boundaries -- using the same multi-scale simulation technique as outlined above.  The segment can be simulated on its own, so as to compare the segment's structural dynamics in an unconstrained conformation, or inside the large bent loop. Or, if the segment in question is a binding site of a certain protein, that protein could be docked to the DNA segment, and the simulation of the two would allow one to study the changes in the protein binding to DNA upon the loop formation.  With the advent of the computational power, even the simulations of the all-atom structures of moderately sized DNA loops, such as the 76~bp loop studied here, would become possible.  For such a simulation, the all-atom loop structure, predicted on the basis of the elastic loop model, can serve as a good starting point.


\subsection{Further developments}

Naturally, before being used in such advanced simulations, the model has to be refined and extensively tested using all the available experimental data. In the present state, we have not even converged to a single preferred set of elastic moduli, as a whole family of values may account for the \lr\ loop bending energies. Refining and adjustment of the model parameters on the basis of other data is at this point imperative for making the suggested universal elastic model fully functional on all levels. Such data can come from the experiments on DNA interaction with other DNA-binding proteins, including topoisomerases~\cite{PULL97,RYBE97}, from the energetics of DNA minicircles~\cite{FRAN85,TRIF91}, and from the analysis of deformations in DNA X-ray structures, following the ideas of Olson {\it et al}~\cite{MATS2002,OLSO98}.

Another necessary adjustment would have to account for the well-known sequence-specificity of DNA properties~\cite{HOGA87,OLSO98,MATS2002}. The bending moduli have to be determined in the sequence-specific fashion, as functions of the arclength $s$, varying according to the DNA sequence at each point of the studied loop (see App.~\ref{sec:appendixd}).  The intrinsic curvatures $\kiot$ and varying intrinsic twist $\omo$ are also known to have a significant effect on the structure and dynamics of some DNA sequences~\cite{MATS2002,KATR97,YANG95} and parameterizing these functions is therefore also important for our model.

Before the parameters are better defined, the model is still good for solving general problems, such as selection between alternative DNA loop topologies or energy estimates within several kT, as discussed in the previous section, but not for more subtle quantitative predictions of the structural and energetical properties of DNA loops.

Several extensions to the model may prove necessary in order to achieve realistic DNA descriptions in certain cases. One obvious extension is including DNA deformability into Eqs.~\refgrandA. DNA is known to be a shearable and extensible molecule, as is evidenced by the deformations observed in the all-atom structures~\footnote{Namely, the deformations of rise, slide, and shift~\cite{MATS2002,OLSO2001,OLSO98,GORI95}.} or micromanipulation experiments~\cite{STRI2000,STRI96,SMIT96}. The DNA deformability can influence both the local and global structure of the modeled DNA, especially in view of the observed coupling between the DNA stretch and twist~\cite{MARK97}. In order to include the DNA deformability in our model, Eqs.~\refgrandA\ have to be modified as outlined in App.~\ref{sec:appendixe}, raising the order of the system \refgrandA\ to 16.

Another vital modification consists in adding a steric repulsion parameter to the other force terms.  In the present study, there was no need for such a term because of its insignificance compared to the DNA self-repulsion.  However, if the DNA interaction with a positively charged object (e.g., the histone core of the nucleosome) is to be described, then the steric repulsion term becomes imperative lest the elastic solutions collapse on the positive charges, causing non-convergence of the iterative process.  The steric repulsion can be described as the van der Waals '6-12' potential and would not be qualitatively different from the electrostatic repulsion introduced in~\eqref{eq:E_field}. Only the electric charges in eq.~\eqref{eq:E_field} have to be replaced by the van der Waals coefficients, the degrees of the denominators have to be changed to 6 or 12, and the inverse screening radius has to be set to zero -- otherwise, the forces of steric repulsion are computed in the same way and through the same iterative algorithm applied to solving the integro-differential equations.

As for the electrostatic self-repulsion, its description in our model can be rendered more realistic by placing the phosphate charges on the outside of the double helix rather than on the centerline as in the present simplified treatment~\eqref{eq:E_field}. Placing the phosphates at the points $\rho_1 \dv(s) + \rho_2 \dt(s)$ of the rod cross-section, where $\rho_1$ and $\rho_2$ are determined by the DNA chemical structure (Fig.~\ref{fig:elrod_dna}), would result in replacing the radii $\rv(s)$ in eq.~\eqref{eq:E_field} by $\vec{R}_{Ph}(s) = \rv(s) + \rho_1 \dv(s) + \rho_2 \dt(s)$. That would make the electric field dependent not only on the radius-vector $\rv(s)$, but also on the the Euler parameters $\qi(s)$ that determine the orientation of the local coordinate frame.  This additional dependence is unlikely to result in any algorithmic difficulties with solving the equations.  Physically, however, moving the phosphate charges away from the centerline means introducing external torques in each cross-section in additional to external forces, and the torques could change the calculated loop structure -- again, unlikely in the general case, but possibly in the case of a close approach of the charged elastic ribbon to itself or any other charged object involved in the model.

As has been noted above, the discussed modeling method produces static, zero-temperature structures of DNA loops. Yet, the entropic contribution to the free energy of different DNA states may sometimes be important. Interestingly, there is a way to estimate the structural entropy of a DNA loop with our method.  One could employ the intrinsic curvature/twist terms and perform statistical sampling by assigning bends and unwinding/overwinding with the energetic penalty between 0 and 1\,kT at random points of the loop, analyzing the resulting changes in the loop structure and energy.  Since the zero temperature structure of the loop is used as the starting point in the iterative calculations of the randomly bent and twisted structures, the iterations should converge much faster than those running from scratch.  Thus the ensemble of thermally excited structures can be generated and used to obtain the properties of the studied DNA loop at a final temperature. The entropy of each looped state can be estimated, for example, through the volume of space swept by all the different structures from the thermal ensemble.

With the modifications, outlined above, the proposed model is likely to describe DNA loops very realistically, yet still be less detailed and computationally much faster than all-atom models. One drawback, however, lies in the non-linear nature of the elastic problems. The solutions to the equations of elasticity are known to exhibit a non-trivial dependence on the problem parameters, for example, the boundary conditions. A latent response to the change of a certain condition can lead to an abrupt change in the shape of the BVP solution, for example, if the point of instability of the latter is achieved.  If the solution changes too abruptly, the iterative procedure will fail due to non-convergence of the BVP solver.  Another reason for the non-convergence is meeting a bifurcation point, as it happened at low salt concentrations with the \UP\ and \OP\ solutions for the 385~bp loop (Sec.~\ref{sec:electro_385}).

Such problems seem to be inherent to our method. If they are encountered, we would recommend to thoroughly analyze the nature of the non-convergence, for example, through monitoring the evolution of the intermediate solutions prior to the non-convergence. In the case of abrupt changes, the ultimate structure can perhaps be guessed, or achieved along a different pathway with an equivalent endpoint that might not be leading the rod through the point of abrupt change (for instance, rotating the end of the elastic rod counterclockwise by $2\pi-\phi$ if a clockwise rotation by $\phi$ is causing problems). In the case of a bifurcation, the bifurcating solutions branches need to be analyzed -- which by itself may provide useful insights into the structural properties and transformations of the studied loop.


\subsection{\Lr\ loops}

To conclude the paper, let us summarize what has been learned with our method about the specific system, the \lr\ and its DNA loop.  For both possible lengths of the loop, it has been shown that the underwound loop structure pointing away from the \lr\ should be predominant under thermodynamic equilibrium conditions, unless other biomolecules interfere.  The predicted structure of the U loop depends only slightly on the salt concentration, although the loop energy exhibits a stronger dependence.  The experimentally observed energy of the loop can be obtained with the right combination of parameters - which, of course, have to be extensively tested with other protein-DNA systems.

Another reason why the predicted structure of the DNA loop has to be treated with caution is the dynamic nature of the \lr-DNA complex.  The X-ray structure~\cite{LEWI96}, on which our conclusions were based, has been obtained without the DNA loop connecting the protein-bound DNA pieces and, therefore, can be presumably changed by the stress of the bent DNA loop.  Yet, our results pave the way for analyzing these changes.  The forces of the protein-DNA interactions, computed with our model, can be applied in a multi-scale simulation of the \lr-DNA complex, as described above.  Such simulation would reveal the equilibrium state of the \lr\ with the bound DNA loop, or at least show the spectrum of structural states visited by the protein during its dynamics with the attached loop.

The results of such dynamics may even change our conclusions about the structure and energetics of the DNA loops created by the \lr. When the boundary conditions change, the O loop can become energetically preferable to the U loop, or even one of the dismissed extraneous solutions may become feasible.  Yet, to make multi-scale simulations possible, that would study the complexes between \lr, or other proteins, and the DNA loops in their dynamic nature, was the driving force behind the development of our coarse-grained DNA model -- which, per se, does not pretend to yield final conclusions about the \lr-DNA complex.


\section{Conclusion}
 
In conclusion, a universal theoretical model of DNA loops has been presented. The model unifies several existing DNA models and provides description of many physical properties of real DNA: anisotropic flexibility, salt-dependent electrostatic self-repulsion, intrinsic twist and curvature -- all the properties being sequence-dependent. The model is applicable to a broad range of problems regarding the interactions of DNA-binding proteins with DNA segments of moderate length (100--1,000\,bp) and can serve as a basis of all-atom or multi-scale simulations of protein-DNA complexes. The application of the model to the \lr-DNA complex revealed a likely structure of the 76~bp and 385~bp DNA loops, created by the protein. The experimentally measured energy of the DNA loop formation is obtainable with a proper set of parameters. The obtained forces of the protein-DNA interactions can be used in a multi-scale simulation of the \lr-DNA complex. Further comparison with experimental data will be beneficial for optimizing the parameters and approximations of the model.

 
\section{Acknowledgments}
This work was supported by grants from the Roy J. Carver Charitable Trust, the National Institute of Health (PHS 5 P41 RR05969), and the National Science Foundation (BIR 94-23827EQ). The figures in this manuscript greatly benefited from the molecular visualization program VMD~\cite{HUMP96}.


\appendix
 
\section{}
\label{sec:appendixa}

The following eleven dimensionless functions constitute the initial simplified solution to the equations~\refgrandA, the planar circular uncharged isotropic rod:
\begin{eqnarray}
\kappa_1 &=& 2 \pi \sin ( \omo s - \psio )\:,\\ 
\kappa_2 &=& 2 \pi \cos ( \omo s - \psio )\:, \\
\omega &=& 0\:, \\
n_3 &=& 0\:, \\
q_1 &=& \sin \pi s \; \sin ((\omo s - \psio) / 2 ) \:,  \\
q_2 &=& \sin \pi s \; \cos ((\omo s - \psio) / 2 )\:, \\
q_3 &=& \cos \pi s \; \sin (\omo s / 2)\:, \\
q_4 &=& \cos \pi s \; \cos (\omo s / 2)\:, \\
x &=& (1/2\pi) (1-\cos 2 \pi s) \; \cos 2 \psio\:, \\
y &=& (1/2\pi) (1-\cos 2 \pi s) \; \sin 2 \psio\:, \\
z &=& (1/2\pi) (\sin 2 \pi s)\:.
\end{eqnarray}
\noindent Here $\psio$ is an arbitrary parameter that determines the angle between the $x$ axis of the LCS and the plane of the elastic rod.

\section{}
\label{sec:appendixb}

According to~\cite{MILL80,HSIE87}, the equilibrium constant of binding of the \lr\ to a single-operator DNA equals about $10^{-11}$~M for O$_1$ and $10^{-9}$~M for O$_3$ at high salt concentration (0.2~M). This results in the free energies of binding $\Delta G_{O_1} = \kT \log K_{O_1} = -25\kT$ and $\Delta G_{O_3} = \kT \log K_{O_3} = -21\kT$. The equilibrium binding constant of the \lr\ to the DNA promoter, containing both O$_1$ and O$_3$ sites, equals $3.4 - 6.2 \times 10^{-12}$~M~\cite{HSIE87}, resulting in the free energy $\Delta G_{O_1-O_3} \approx -26\kT$.  This results in the free energy of formation of the 76~bp DNA loop $\Delta G_{loop} = \Delta G_{O_1-O_3} - \Delta G_{O_1} - \Delta G_{O_3} = 20\kT$.

\section{}
\label{sec:appendixc}

In order to derive~\eqref{eq:bendmod_tw}, let us consider a short section of a tightly twisted rod with no intrinsic curvature.  Then the total curvature equals $K(s) = \sqrt{ K_1(s)^2 + K_2(s)^2} = \sqrt{ \ko(s)^2 + \kt(s)^2}$. The bending energy of the section is:
\begin{equation}
\begin{split}
U_{\kappa} = & \int_{s_1}^{s_2} \left( \frac{A_1}{2} \ko^2 + \frac{A_2}{2} \kt^2 \right) ds = \\
& \int_{s_1}^{s_2} \left( \frac{A_1+A_2}{4}(\ko^2+\kt^2) + \frac{A_1-A_2}{4}(\ko^2-\kt^2) \right) ds = \\
& \frac{A_1+A_2}{4} \int_{s_1}^{s_2} K^2 ds + \frac{A_1-A_2}{4} \int_{s_1}^{s_2} K^2 \cos 2 \Gamma ds \;,
\end{split}
\label{eq:Uk_terms}
\end{equation}
\noindent where $\Gamma(s)$ is the angle between the binormal $\vec{b}$ and $\dv$ (Fig.~\ref{fig:elrod}\,b). A simple textbook calculation using Frenet formul\ae~\cite{KREY91} yields 

\begin{equation}
\begin{split}
\Gamma(s) = & \int_{s_1}^{s} ( \tau(\sP) - \Omega(\sP) ) d\sP = \\
 & \int_{s_1}^{s} ( \tau(\sP) - \omega(\sP) - \omo(\sP) ) d\sP = \\
 & s <\tau>_s - s <\omega>_s - s <\omo>_s \:,
\end{split}
\label{eq:gam_om_tau}
\end{equation}
\noindent where $\tau(s) = - \ddot{\rv}(s) \cdot \dot{\vec{b}}(s) / K(s)$ is the torsion of the centerline $\rv(s)$ and the angular brackets $<...>_s$ denote the average over the interval $(s_1,s)$.

Since the rod is tightly twisted, the term $s <\omo>_s$ grows much faster with $s$ than both other terms in \eqref{eq:gam_om_tau} and $K(s)$. Therefore, the second term in \eqref{eq:gam_om_tau} is an integral of a fast oscillating function and as such is much smaller than the first term. Accordingly,
\begin{equation}
U_{\kappa} \approx \frac{A_1+A_2}{4} \int_{s_1}^{s_2} K^2 ds\;,
\end{equation}
\noindent and, since the approximation of a single bending modulus assumes
\begin{equation}
U_{\kappa} = \int_{s_1}^{s_2} \frac{A}{2} K^2 ds\;,
\end{equation}
\noindent we arrive at~\eqref{eq:bendmod_tw}.

\section{}
\label{sec:appendixd}

The sequence-dependent parameters could be related to the arclength of the rod in several possible ways.  The simplest would be a step-wise assignment of the parameters. At the points, corresponding to each DNA step between two neighboring base pairs, the parameters (elastic moduli $\alpha$, $\beta$, and $\gamma$, intrinsic curvature $\kiot$ and twist $\omo$) would adopt the values correspondent to that DNA step. Connecting those points with smooth functions (for example, spline-based ones) would result in the desired parameter setup $\alpha(s)$, $\beta(s)$, etc., for the particular loop.

A slightly modified approach could keep the DNA step-based parameters constant in a certain area in the middle between the base pairs and limit the zone of a smooth transition from value to value to a certain width $\delta$. Then, the sequence-based parameter functions between the DNA steps to and from the base pair located at the point $s_i$ would look like:
\begin{equation}
\label{eq:elast_seqsp}
\begin{split}
\alpha(s) = & \;\alpha_i\:, \quad\quad {\rm if}\ s_i - h/2 \leq s < s_i - \delta/2 \\
\alpha(s) = & \;\halfof{\alpha_{i+1}+\alpha_i} + \halfof{\alpha_{i+1}-\alpha_i} \sin \frac{\pi}{\delta} (s-s_i) \:, \\
 & \quad\quad\quad\quad  {\rm if}\ s_i - \delta/2 \leq s \leq s_i + \delta/2 \\
\alpha(s) = & \; \alpha_{i+1}\:, \quad {\rm if}\ s_i + \delta/2 < s + h/2 \leq s_i\;.
\end{split}
\end{equation}
\noindent Certainly, another smoothly differentiable function can be used in the transition zone $s_i-\delta/2 \leq s \leq s_i+\delta/2$ instead of the sine.

Finally, a more complicated approach can relate the parameters to a longer DNA sequence surrounding each point $s$ rather than to only three neighboring base pairs defining the two adjacent DNA steps. Such approach would be more realistic, as the experimental and simulation data indicate~\cite{HOGA87,OLSO98}. In that case, the model parameter functions would have to depend on multiple base pairs neighboring each point $s$, for example, as described in~\cite{HOGA87}.

\section{}
\label{sec:appendixe}

The DNA deformability can be described in our model by additional three variables, combined into the shift vector $\veps(s)$. Its components $\epsilon_{1,2}$ are the amount of shear in the two principal directions, and the component $\epsilon_3$ is the amount of extension along the normal $\dth$~\cite{SHI95,WEST97}.  The vector of shift $\veps$ and the elastic force $\vec{N}$ are linearly related to each other, similarly to the vector of curvature $\vec{k}$ and the torque $\vec{M}$ (cf eq.~\eqref{eq:ber_eul}):
\begin{equation}
\label{eq:n_ber_eul}
\vec{N}(s) = B_1 \epsilon_1 \dv + B_2 \epsilon_2 \dt + D \epsilon_3 \dth \;,
\end{equation}
\noindent where $B_{1,2}$ are the shear moduli in the two principal directions, and $D$ is the extension modulus of DNA. Those parameters would also have to be determined from and extensively verified against the experimental data, e.g. those presented in~\cite{OLSO98,GORI95}.

Thus introduced, deformability changes eq.~\eqref{eq:inext} into
\begin{equation}
\label{eq:tang_ext}
\dot{\vec{r}} = \veps + \dth \;,
\end{equation}
\noindent propagating into eq.~\eqref{eq:tor_eq}.  Finally, the systems \refgrand, \refgrandA\ become of 16-th rather than 13-th order, but could be similarly solved by the continuation method.

\section{}
\label{sec:appendixf}

The following algorithm has been used in this work to build all-atom DNA structures on the basis of coarse-grained elastic rod solutions. First, the all-atom structures of the base pairs, with the sugar phosphate backbone groups attached, were built in the idealized B-conformations using Quanta~\cite{QUAN88}.  Second, the chosen elastic rod solution was used to obtain local coordinate frames \dframes{s_{bp,i}} from $q_i(s_{bp,i})$~\cite{WHIT60} at each point $s_{bp,i} = ih/L$ corresponding to the location of each base pair of the loop along the centerline of the DNA helix. Third, the all-atom base pairs were centered at the points $\rv(s_{bp,i})$ and aligned with the coordinate frames \dframes{s_{bp,i}} as illustrated in Fig.~\ref{fig:elrod_dna}. Fourth, the sugar phosphate groups were connected with each other and the \lr-bound DNA segments by phosphodiester bonds. Fifth, two 50-step minimization rounds with X-PLOR~\cite{BRUN92b} using CHARM22 force field~\cite{MACK95} were performed in order to relieve bad interatomic contacts and chemical group conformations (bonds, angles, dihedral angles) resulting from this idealized placement, especially for the DNA backbone. The resulting all-atom structure, while still stressed at certain points and overidealized at others, presents a reasonable starting point for any all-atom or multi-scale simulations, as described in Sec.~\ref{sec:discussionB}.



\end{document}